\documentclass[amsmath,amssymb,showpacs,twocolumn,superscriptaddress]{revtex4-1}
\usepackage{graphicx}
\usepackage{tabularx}
\usepackage{color}
\usepackage{amsmath}
\usepackage{comment}
\newcommand{\be}{\begin{equation}}
\newcommand{\ee}{\end{equation}}
\newcommand{\bea}{\begin{eqnarray}}
\newcommand{\eea}{\end{eqnarray}}
\usepackage{dcolumn}
\usepackage{hyperref}
\usepackage{bm}
\usepackage{epsf}
\usepackage{subfigure}
\usepackage{epstopdf}%
\usepackage{natbib}
\usepackage{amsfonts}

\DeclareMathOperator{\trace}{Tr}

\begin{document}
\title{
Path-integral methodology and simulations of quantum thermal transport:  \\
Full counting statistics approach}

\author{Michael Kilgour}
\affiliation{Chemical Physics Theory Group, Department of Chemistry,
University of Toronto, 80 Saint George St., Toronto, Ontario, Canada M5S 3H6}

\author{Bijay Kumar Agarwalla}
\affiliation{Department of Physics, Indian Institute of Science Education and Research, Dr. Homi Bhaba Road, Pune, India}

\author{Dvira Segal}
\affiliation{Chemical Physics Theory Group, Department of Chemistry,
University of Toronto, 80 Saint George St., Toronto, Ontario, Canada M5S 3H6}
\affiliation{Centre for Quantum Information and Quantum Control,
University of Toronto, 80 Saint George St., Toronto, Ontario, Canada M5S 3H6}

\date{\today}
\begin{abstract}
We develop and test a computational framework to study heat exchange in  
interacting, nonequilibrium open quantum systems.
Our iterative full counting statistics path integral (iFCSPI) approach extends a previously well-established influence functional path integral method, by going beyond reduced system dynamics to provide the cumulant generating function of heat exchange.
The method is straightforward; we implement it for the nonequilibrium spin boson model to calculate transient and long-time observables, focusing on the steady-state heat current flowing through the system under a temperature difference.
Results are compared to perturbative treatments and demonstrate good agreement in the appropriate limits.
The challenge of converging nonequilibrium quantities, currents and high order cumulants, is discussed in detail.
The iFCSPI, a numerically exact technique, naturally captures strong system-bath coupling and non-Markovian effects of the environment. 
As such, it is a promising tool for probing fundamental questions in quantum transport and quantum thermodynamics.
\end{abstract}

\maketitle

\section{Introduction}
\label{Sec-intro}

Understanding the transfer of energy in nanoscale systems 
is of central importance for the design of energy efficient devices 
\cite{Pop,ReddyJCP,ARPCSegal}, and for uncovering fundamental physical bounds on the rates of computation and information flow  \cite{Pendry,Blencowe,Markov,Ed}.  
How do particles and energy travel through nanostructures, quantum dots, small or large organic and biological molecules?
At the nanoscale, quantum effects may play an instrumental role in 
the management of waste heat and the function of electronic, thermal and thermoelectric devices.
Furthermore, understanding quantum heat flow is a critical step
for further developments in quantum thermodynamics \cite{Millen2016,Vinjanampathy2016,Alicki2018}.

In this paper, we focus on the problem of quantum heat transfer in nanojunctions.
Beyond the average heat current,  
fluctuations are of significant interest, in particular at the nanoscale when they 
are substantial \cite{Esposito-review,Hanggi-review}.
A full-counting statistics (FCS) analysis provides the
probability distribution function $P_t(Q)$ of the transferred heat $Q$ within a time interval $t$.
Obtaining this function in interacting systems is a formidable task, achieved through the
development of e.g. perturbative \cite{renJ,Nicolin2011,Nicolin2011b,bijay2012, Friedman2018,Agarwalla2017a,Ren,Ren2,Wang2013}
and numerically-exact \cite{brandes, Weiss,Weiss2,ShiHEOM} treatments.
FCS calculations not only hand over the cumulants of heat exchange,
but allow one to test  fundamental relations, such as the steady state heat exchange
fluctuation symmetry \cite{GC-sym,Esposito-review,Hanggi-review}.

The non-equilibrium spin-boson (NESB) model generalizes the
dissipative spin-boson model \cite{Weiss-book} to include two or more heat baths. As such, it
serves as a test-bed to explore the fundamentals of quantum heat 
flow in anharmonic nanojunctions \cite{segal-PRL,segal-QME,nazim}.
In the NESB model, the spin represents an anharmonic mode, constructed from a truncated harmonic spectrum,
and the heat baths are described by a collection of harmonic oscillators (phonons).
The NESB model is an extremely rich platform for studying nonlinear
effects in quantum transport \cite{ARPCSegal},
including the operation of thermal diodes \cite{segal-PRL, segal-QME}, transistors \cite{Rentrans},  
and heat machines \cite{segal-pump,segal-stochastic,renJ}.
To simulate the transients and steady state behavior of heat exchange in the NESB model, 
efforts have been made to generalize 
methodologies originally developed for time evolving the reduced density matrix. 
A partial list includes
perturbative (Born-Markov) quantum master equation tools
\cite{segal-PRL, segal-QME,Ojanen-QME},
the noninteracting blip approximation (NIBA) \cite{segal-PRL, segal-QME, Nicolin2011,Nicolin2011},
the nonequilibrium polaron-transformed Redfield equation \cite{Ren, Ren2}, 
Green's function methods \cite{ThossJCP,Juzar14,WuEPL,Liu17,Agarwalla2017a}, 
wavefunction approaches \cite{num-multi}, hierarchical equations of motion \cite{Tanimura15,Tanimura16,brandes},
path integral methodologies \cite{SegalPI13,Saito13, SaitoNJP,Yamamoto18,Weiss}, and mixed quantum-classical equations
\cite{Junjie18}.

In this work, we present and apply a new tool for the study 
of the full counting statistics (FCS) of heat exchange in nanojunctions.
Our numerically exact path integral method, which we name the
``iterative full counting statistics path integral'' (iFCSPI) method,
provides the steady state cumulants of heat exchange, 
specifically, the averaged heat current and its noise.
Focusing on the NESB model we perform simulations of the heat current beyond the
weak system-bath coupling and the nonadiabatic (weak tunneling) transport limits.

The iFCSPI is built on the 
influence functional path integral representation of quantum dissipative dynamics \cite{Feynman1963},
with the influence functional modified to include counting information.
The implementation of the iFCSPI is based on combining 
(i) the two-time measurement protocol, 
which rigorously builds the characteristic function for heat exchange, and
(ii) the iterative quasi-adiabatic influence functional path integral approach (iQuAPI) of Makri and Makarov \cite{Makri1995,Makri1995a}, which was originally developed for iteratively time-evolving the reduced density matrix. 
Therefore, similarly to iQuAPI, the convergence of iFCSPI to the exact limit relies on
the existence of a well-defined decorrelation time of the nonlocal influence functional.

Like other iterative influence functional approaches, the iFCSPI relies on discretization of the system and time coordinates.
The standard iQuAPI approach can be difficult to converge depending on the system 
and bath parameters, though significant progress with improved algorithms 
based on e.g. path filtering \cite{Lambert2012} or alternative construction of the propagator
has been made in recent years \cite{MakriQC1,MakriQC2,Makri2017,Strathearn2018}. 
Nevertheless, to establish our method and understand its strength and limitations, we here 
implement iFCSPI based on the elementary iQuAPI algorithm \cite{Makri1995,Makri1995a}. 

The paper is organized as follows. 
We introduce the NESB model and the FCS formalism in Sec. \ref{Sec-NESB}. 
We describe the iFCSPI algorithm in Sec. \ref{Sec-Algorithm},
with technical details set aside for Appendices A and B.
Simulation results are presented in Sec. \ref{Sec-results};
Convergence and error analysis are discussed in Appendix C.
We conclude in Sec. \ref{Sec-Summ}. 

\section{Full counting statistics for heat exchange}
\label{Sec-NESB}

\subsection{The nonequilibrium spin-boson model}

The iFCSPI approach is presented here in the language of the Feynman-Vernon (FV) influence functional \cite{Feynman1963}.
The main assumptions underlying this approach are that the environment is composed of 
harmonic oscillators, and that
the system-bath interaction Hamiltonian is a linear function in the baths' displacements.
As such, the method naturally fits to describe the Caldeira Leggett and the spin-boson models \cite{Weiss-book}. 
Nevertheless, the iFCSPI method can be extended beyond this construction
to setups missing an analytic form for the influence functional 
and including correlations beyond pairwise interactions \cite{Makri1999,Segal2010}.
For simplicity, here we implement the iFCSPI within the minimal NESB model.

The Feynman-Vernon path integral approach has been 
recently used to study work statistics \cite{QuanPRL} and heat statistics \cite{QuanPRE, Weiss,Weiss2} 
in the Caldeira-Leggett and the spin-boson models. However, these studies were focused on describing the analytical properties
of the path integral formula for heat exchange and work (e.g. proving the Jarzynski's equality for work fluctuations), 
while our focus here is on the presentation of a feasible numerical implementation to compute heat transfer cumulants.

Let us now describe the NESB model. The total Hamiltonian consists of a spin system $\hat{H}_S$,  
which is coupled via $\hat{H}_{S\alpha}$ to two reservoirs $\hat{H}_{\alpha}$ ($\alpha=L,R$),  
\begin{equation}
\hat{H}= \hat{H}_S + \hat{H}_L + \hat{H}_R + \hat{H}_{SL} + \hat{H}_{SR}.
\end{equation}
The $L$ and $R$ heat baths are assumed to be prepared 
at thermal equilibrium at the inverse temperatures $\beta_L$ and $\beta_R$, respectively. 
In the long time limit, a nonequilibrium steady state in reached 
when the thermal baths are set at different temperatures.
In the NESB model, the system comprises a single spin,
\begin{equation}
\hat{H}_S= \frac{\omega_0}{2}\hat{\sigma}_z+\frac{\Delta}{2}\hat{\sigma}_x,
\end{equation}
with $\hat \sigma_{x,z}$ as the Pauli matrices, 
$\omega_0$ the spin splitting, and $\Delta$  the tunneling energy. 
For simplicity, in what follows we take $\omega_0=0$,
therefore $\Delta$ becomes the level splitting in the system eigenbasis. 
The bath plus system-bath interaction,
collectively, the ``environmental Hamiltonian'' are given by
\bea
\hat{H}_{env}=\sum_{\alpha=L,R}\left(\hat{H}_\alpha+\hat{H}_{S\alpha}\right). 
\eea
We model the baths by collections of independent harmonic oscillators. 
These oscillators are bilinearly coupled to the spin, 
$\hat{H}_{S\alpha}=\hat{S}\otimes\hat{B}_\alpha$, 
with $\hat S$ as the system operator (spin polarization $\hat{\sigma}_z$)
and $\hat B_{\alpha}=\sum_k \lambda_{\alpha,k}(\hat b_{\alpha,k}^{\dagger}+\hat b_{\alpha,k})$ 
given in terms of 
the displacements of the baths' normal modes. 
Altogether, the environmental Hamiltonian is given by
\bea
\hat{H}_{env}=
\sum_{\alpha,k}\bigg[\omega_{\alpha,k}\hat b^\dagger_{\alpha,k}\hat b_{\alpha,k}
+
\lambda_{\alpha,k}(\hat b^\dagger_{\alpha,k}+\hat b_{\alpha,k})  \hat{\sigma}_z 
 \bigg].
\eea
Here, $\hat b_{\alpha,k}^\dagger$ ($\hat b_{\alpha,k}$) are bosonic creation  (annihilation) operators for modes 
of frequency  $\omega_{\alpha,k}$ in the $\alpha$ bath. The system-bath interaction 
is characterized by a spectral density function,
\bea
g_\alpha(\omega)=\pi \sum_k \lambda_{\alpha,k}^2\delta(\omega-\omega_{\alpha,k}). 
\eea
In simulations we employ an Ohmic function of the form 
$g_\alpha(\omega)=\gamma_\alpha \omega e^{-\omega/\omega_{c}}$. 
The method is general for other physical spectral densities, 
though the convergence characteristics depend intimately on this choice.
The dimensionless system-bath coupling parameter $\gamma$ 
is related to the dimensionless Kondo parameter $K$, $\gamma=\pi K/2$ \cite{Weiss-book}. 
$\omega_c$ is the cutoff frequency of the baths, 
taken here as identical between hot and cold reservoirs. 

\begin{figure}[htpb]
{\includegraphics[scale=.5]{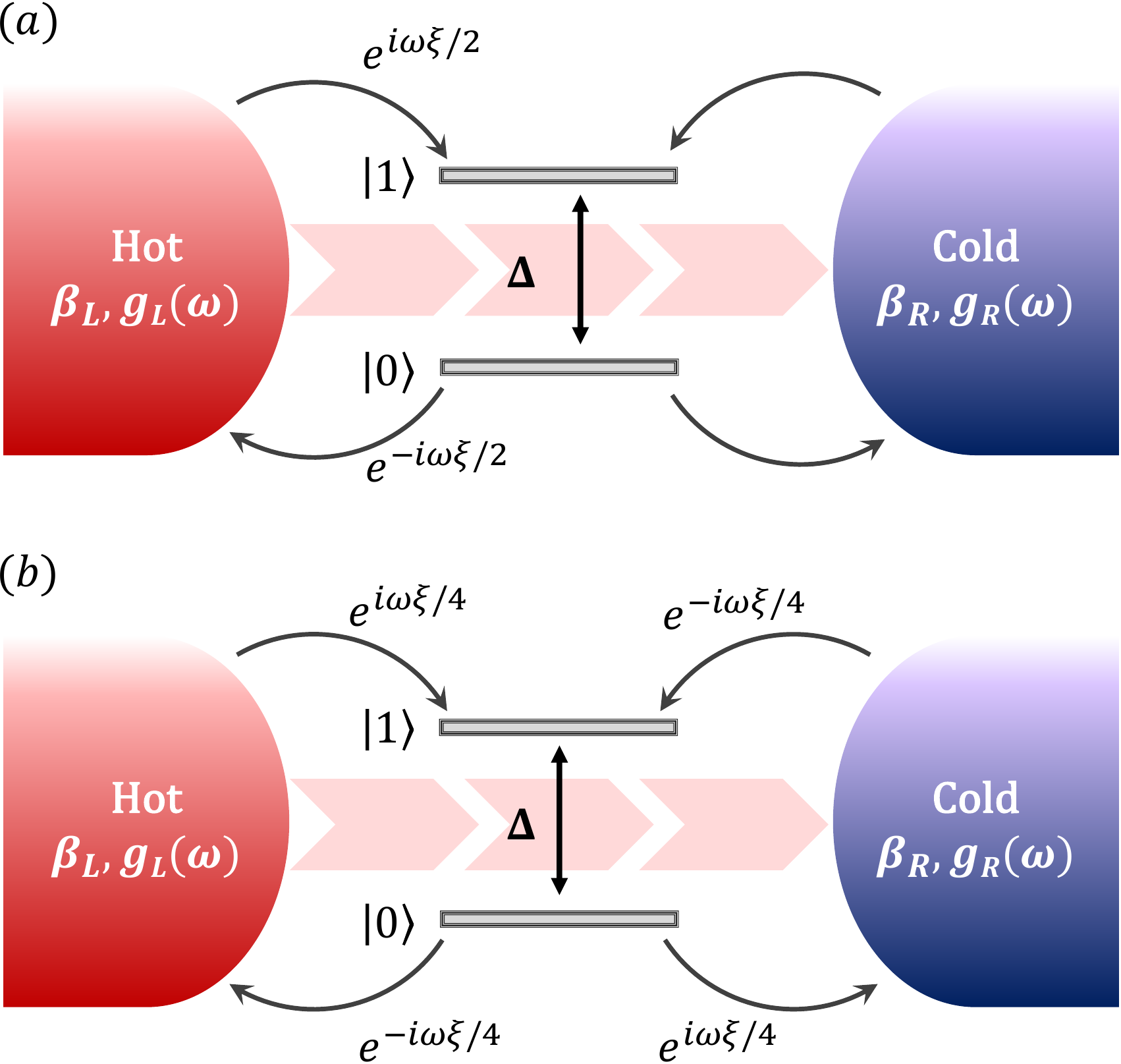}} 
\caption{Scheme of the NESB nanojunction, while illustrating with curvy arrows the counting process with the phase factors decorating the system-bath interaction, see Eqs. (\ref{eq:QL})-(\ref{eq:Uxi4}). The expansive arrows display the direction of the net heat flow in steady state.
(a) Counting system-bath heat exchange only at the left contact.
(b) Counting heat exchange in a symmetrical manner.
Model parameters include $\Delta$ as the spin spacing in the energy basis. The baths
are characterized by spectral functions 
$g_{\alpha}(\omega)$  and an inverse temperature $\beta_{\alpha}$.}
\label{fig:scheme}
\end{figure}
\subsection{Cumulant generating function for heat transfer}

To glean information about transport statistics 
we compute the generating function of heat exchange. 
First, we define the energy current operator 
as the rate of change of energy in one of the reservoirs, 
$\hat J_{Q,\alpha}(t)= - {d\hat{H}^{H}_\alpha(t)}/{dt}$. 
Operators are written in the Heisenberg representation 
and evolve with respect to the total Hamiltonian $\hat{H}$. 
Therefore, the total energy change in the $L$ bath within the time interval $t_0=0$ to $t$ 
is given by the integrated current
\bea
Q_{L}(t,t_0=0)&=& \int_{t_0=0}^{t} \hat J_{Q,L}(t') dt'
\nonumber\\
&=& \hat{H}_{L}(0)- \hat{H}_{L}^{H}(t).
\label{eq:QL}
\eea
We define energy leaving the left bath towards the system as positive.
Note that since the system is not driven by an external field, 
the energy exchanged is thermal in nature, i.e. heat.
Following this definition of heat exchange, we write down its moment generating function 
based on the two-time measurement protocol \cite{Esposito-review, Hanggi-review,bijay2012},
\bea
{\cal Z}(t,\xi) \equiv \Big \langle e^{i\xi \hat{H}_L}e^{-i \xi \hat{H}_L^H(t)} \Big \rangle.
\label{eq:Z_asym}
\eea
Here, $\xi$ is referred to as the ``counting field'' as it tracks the energy transferred to the bath.
The expectation value is evaluated with respect to the total initial density matrix, which is assumed to
be system-bath factorized,  $\rho_0=\sigma_0\otimes\rho_B$.
$\sigma_0$ is the density matrix of the spin system at $t=0$. 
The state of the two baths is initially factorized as well, $\rho_B=\rho_L\otimes\rho_R$, 
each prepared at a thermal equilibrium state, 
$\rho_\alpha=\frac{e^{-\beta_\alpha\hat{H}_\alpha}}{{\rm Tr} [e^{-\beta_\alpha\hat{H}_\alpha}]}$.

In the long time limit, the cumulant generating function (CGF) for heat exchange is defined as
\bea
{\cal G}(\xi) &\equiv& \lim_{t \to \infty} \frac{1}{t}{\rm ln} \,{\cal Z}(t,\xi)
\nonumber\\
&=&\lim_{t \to \infty} \frac{1}{t}\sum_{n=1}^{\infty} \frac{(i\xi)^n}{n!}\langle\langle Q_L^n(t,0)\rangle \rangle,
\label{eq:CGF}
\eea
with $\langle \langle Q_L^n(t,0) \rangle \rangle$ standing for the cumulants for heat exchange between the system
and the $L$ bath within the interval $[0,t]$. Explicitly,
\bea
{\cal G}(\xi) 
=(i\xi)\frac{\langle Q_L(t,0)\rangle}{t}\Big|_{t\to \infty}
+\frac{(i\xi)^2}{2} \frac{\langle\langle Q_L^2(t,0)\rangle \rangle}{t}\Big|_{t\to \infty} + ...,
\nonumber\\
\label{eq:CGFE}
\eea
where we calculate cumulant in the long time limit when the CGF reaches a constant value.
To find the steady-state heat current, we take the first derivative of the CGF with respect to the counting parameter $\xi$,
\bea
\langle J_{Q}\rangle=\frac{d{\cal G}(\xi)}{d(i\xi)}\Bigg\rvert_{\xi=0}.
\label{eq:J}
\eea
Higher cumulants are computed as higher derivatives.

The counting field has the effect of ``dressing'' the forward and backward time evolution operators
$e^{\mp i\hat{H}t}$.
This can be observed from Eq. (\ref{eq:Z_asym}) after making some rearrangements,
\bea
{\cal Z}(t,\xi)&=& 
\Big\langle e^{i\xi \hat{H}_L}\hat{U}^\dagger(t)e^{-i\xi \hat{H}_L}\hat{U}(t)\Big\rangle
\nonumber\\
&=& {\rm Tr} \Big[ e^{-\frac{i\xi}{2} \hat{H}_L} \hat{U}(t) e^ { \frac{i\xi}{2} \hat{H}_L}
\rho_0  e^{  \frac{i\xi}{2} \hat{H}_L}  \hat{U}^\dagger(t)  e^{  -\frac{i\xi}{2} \hat{H}_L} 
\Big]
\nonumber\\
&=&{\rm Tr}\Big[ \hat{U}_{-\xi}(t)\rho_0 \hat{U}^\dagger_\xi(t)\Big]
\nonumber\\
&=&{\rm Tr}_S\big[{\rm Tr}_B[\rho^{\xi}(t)]\big]
\label{eq:Z_asym2}
\eea
Throughout this work, $\xi$'s are assigned to operators dressed by counting fields. 
In the last line we define $\rho^\xi(t)$, which closely resembles the standard density operator,
but evolving with modified, counting field dependent propagators. Similarly,
${\rm Tr}_B [\rho^{\xi}(t)]$ can be regarded as the counting field dependent reduced density matrix.
The dressing of the propagators is defined as
\bea
\hat{U}_{-\xi}(t)&=&e^{-\frac{i\xi}{2} \hat{H}_L} \hat{U}(t)e^{\frac{i\xi}{2} \hat{H}_L} = e^{-i\hat{H}_{-\xi} t}, 
\nonumber\\
\hat{U}^\dagger_{\xi}(t)&=&e^{\frac{i\xi}{2} \hat{H}_L} \hat{U}^\dagger(t)e^{-\frac{i\xi}{2} \hat{H}_L} = e^{i\hat{H}_{\xi} t},
\eea
and we used the fact that 
$\hat A e^{i\hat H t}\hat A^{\dagger}=e^{i\hat A\hat H\hat A^{\dagger}}$ for a unitary $\hat A$ to define the dressed Hamiltonians $H_{\pm \xi}$ above.
This transformation
impacts only the system-bath interaction part of the Hamiltonian. 
For example, the forward time evolution operator 
$e^{-i\hat{H}_{-\xi} t}$ is given in terms of 
\bea
\hat{H}_{-\xi}= \hat{H}_S + \hat{H}_L + \hat{H}_R + \hat{H}^{-\xi}_{SL} + \hat{H}_{SR},
\eea
where the transformed system-bath Hamiltonian is 
\bea
\hat{H}^{-\xi}_{SL}=
\hat{\sigma}_z \sum_k 
\lambda_{L,k}\big(\hat b^\dagger_{L,k}e^{-i\omega_{L,k}\xi/2}+\hat b_{L,k}e^{i\omega_{L,k}\xi/2}\big).
\nonumber\\
\label{eq:bath_shift}
\eea
The counting field therefore enters as phase factors inside the interaction Hamiltonian. 
For a schematic representation of the model and the counting process, see Fig. \ref{fig:scheme}(a).
From here, one can follow the regular derivation of the Feynman-Vernon influence functional, 
and obtain the counting-field dependent analogue.
In Appendix A we show a detailed derivation based on a cumulant expansion. 
In Appendix B we discuss the exact derivation of the influence functional following Ref. \cite{Weiss}.

Back to Eq. (\ref{eq:Z_asym2}), in order to extract information on the heat current,
we need to time evolve the counting field dependent reduced density matrix, 
obtain the generating function ${\cal Z}(t,\xi)$
and compute the cumulant generating function ${\cal G}(\xi)$ 
via a numerical derivative in the long time limit. 
To find the steady-state heat current, one further needs to take the 
derivative of the CGF with respect to the counting parameter, see Eq. (\ref{eq:J}).

To facilitate error cancellation as explained in Appendix C, 
it is necessary in this implementation to apply the counting field in 
a symmetrized manner with respect to the two baths \cite{thossC};
for a schematic representation, see Fig. \ref{fig:scheme}(b). 
Therefore, we repeat steps (\ref{eq:QL})-(\ref{eq:Z_asym2}): 
We define the symmetrized heat exchange in the interval $[0,t]$,
\bea
Q(t)=\frac{1}{2}\Big[\hat{H}_L(0)-\hat{H}_L^H(t)\Big]-\frac{1}{2}\Big[\hat{H}_R(0)-\hat{H}_R^H(t)\Big],
\nonumber\\
\label{eq:Q_sym}
\eea
and the generating function of heat exchange 
\bea
{\cal Z}(t,\xi)= \Big \langle e^{i\xi (\hat{H}_L-\hat H_R)/2}e^{-i \xi \left[\hat{H}_L^H(t)-H_R^H(t)\right]/2}\Big \rangle.
\label{eq:Z_sym}
\eea
These definitions again allow us to arrive at 
Eq. (\ref{eq:Z_asym2}), but with  symmetrized time evolution operators
\bea
\hat U_{-\xi}(t) = e^{-i \xi(\hat{H}_L-\hat{H}_R)/4}\hat{U}(t)e^{i \xi (\hat{H}_L - \hat{H}_R)/4}
\eea
and
\bea
\hat U_{\xi}^{\dagger}(t) = e^{i \xi(\hat{H}_L-\hat{H}_R)/4}\hat{U}^\dagger(t)e^{-i \xi (\hat{H}_L - \hat{H}_R)/4}.
\label{eq:Uxi4}
\eea
%
Note that in the symmetrized representation, 
the sign of the counting field is flipped between the two baths, and its magnitude is halved.

\section{FCS Path Integral}
\label{Sec-Algorithm}

A variety of approaches have been used to compute FCS for quantum 
heat transport problems using the two-time measurement protocol, 
including Green's functions \cite{Agarwalla2017a,Wang2013}, 
quantum master equations\cite{Nicolin2011,Friedman2018,Ren,Ren2}, hierarchical equations of motion \cite{brandes} and mixed quantum-classical methods 
\cite{Junjie18}.
Here, we implement the two-time measurement protocol
within the machinery of the iQuAPI 
scheme developed by Makri and Makarov \cite{Makri1995,Makri1995a}.

\subsection{The influence functional}
We start by presenting the influence functional path integral.
We discretize the time evolution, then break the time evolution operators into ``free'' (system) and ``environmental'' parts
using the symmetrized Trotter decomposition formula.  For example,
the forward time evolution is organized as
\bea
e^{-i \hat{H}_{-\xi} t} &=& \Big(e^{-i \hat{H}_{-\xi} \delta t} \Big)^N 
\nonumber\\
&=&\lim_{N \to \infty}\Big(e^{-i\hat{H}^{-\xi}_{env} \delta t/2} \, 
e^{-i \hat{H}_S \delta t}\, e^{-i \hat{H}^{-\xi}_{env} \delta t/2}\Big)^N,
\nonumber\\
\eea
where $t=N \delta t$, with $N$ as the number of time steps of duration $\delta t$. 
We note again that the sign of the counting field flips in the backward propagator.
%
Following the usual derivation \cite{Makri1995},
we insert resolutions of the system's identity $\hat I=\sum_{s}|s\rangle \langle s|$,
 between each of the short time propagators, 
and take the trace over the two baths' degrees of freedom. 
We thereby obtain the generating function in a path integral form,
\begin{widetext}
\bea
{\cal Z}(t,\xi) &=&\sum_{\{s^{\pm}\}} \langle s_N | e^{-i\hat{H}_S\delta t}|s^+_{N-1}\rangle \cdots \langle s^+_1 | e^{-i\hat{H}_S\delta t}|s^+_{0}\rangle  \langle s_0^+| \sigma_0 |s_0^-\rangle \langle s^-_0 | e^{i\hat{H}_S\delta t}|s^-_{1}\rangle\cdots \langle s^-_{N-1} | e^{i\hat{H}_S\delta t}|s_{N}\rangle
\nonumber\\
&\times& {\rm Tr}_B\Big[e^{-i\hat{H}_{env}^{-\xi}(s_N)\delta t /2}e^{-i\hat{H}_{env}^{-\xi}(s^+_{N-1})\delta t}\cdots e^{-i\hat{H}_{env}^{-\xi}(s^+_{0})\delta t/2}\rho_B e^{i\hat{H}_{env}^{\xi}(s^-_{0})\delta t/2}\cdots e^{i\hat{H}_{env}^{\xi}(s^-_{N-1})\delta t}e^{i\hat{H}_{env}^{\xi}(s_N)\delta t /2}\Big]
\nonumber\\
&=&
\sum_{\{s^{\pm}\}} K(s_0^{\pm}, s_{1}^{\pm})\cdots K(s_{N-2}^{\pm}, s_{N-1}^{\pm})K(s_{N-1}^{\pm}, s_N) 
\langle s_0^+| \sigma_0 |s_0^-\rangle
 \, I^{\xi/2}_L (s_0^{\pm}, s_1^{\pm}, \cdots, s_{N})\, I^{-\xi/2}_R (s_0^{\pm}, s_1^{\pm}, \cdots, s_{N})
\label{eq:PI}
\eea
\end{widetext}
Here, $+$  ($-$) refers to the forward (backward) time evolution. 
Since we are evaluating a trace over the system, 
$s_N^+=s_N^-=s_N$ for the outer sum.

The free evolution is contained in 
$K(s_j^{\pm} ,s_{j+1}^{\pm}) = \langle s_{j+1}^{+} | e^{-i \hat{H}_S \delta t} | s_{j}^{+} \rangle  \, \langle s_{j}^{-} | e^{i \hat{H}_S\delta t} | s_{j+1}^{-} \rangle$. 
The influence functionals,
 $I^{\xi/2}_\alpha(\{s^\pm\})$ carry the details of the environment, its interaction with the system,
and the counting field. 
Since the baths are independent, their respective influence 
functionals factorize to a simple product, see Appendix A.
The counting parameter appears in both influence 
functionals since we used the symmetrized definition of the FCS.

An analytic form for the influence functional can be derived
assuming that: (i) The baths include harmonic oscillators.
(ii) Each  bath is prepared at thermal equilibrium. 
(iii) The system-bath coupling is linear in the bath coordinates.
One obtains then the FV influence functional, which is dressed here by the counting parameters \cite{Weiss}, 
%
%
\begin{widetext}
\begin{equation}
\begin{split}
I_\alpha^\xi(\{s^\pm\})=\exp\Bigg[-\int_0^td\tau\int_0^\tau  d\tau'\big[&s^+(\tau)s^+(\tau')\eta_\alpha(\tau-\tau')-s^-(\tau)s^+(\tau')\eta_\alpha(\tau-\tau'+\xi)\\
-&s^+(\tau)s^-(\tau')\eta_\alpha(\tau'-\tau+\xi)+s^-(\tau)s^-(\tau')\eta_\alpha(\tau'-\tau)\big]\Bigg].
\label{eq:IF}
\end{split}
\end{equation}
\end{widetext}
Recall: In the symmetrized implementation $\xi$ is chopped by a factor of 2, and it further receives a negative sign for the 
$R$ terminal.
In Appendix A we include a derivation of the counting field dependent influence functional
based on a weak-coupling expansion, which is valid beyond the linear coupling-harmonic bath model.
Here, $\eta_{\alpha}(t)$ is the autocorrelation function of the $\alpha$ bath's operators that are coupled to the spin,
%
\begin{equation}
\eta_{\alpha}(t+\xi) = \langle \hat{B}^{\xi}_{\alpha}(t) \hat{B}^{-\xi}_{\alpha}(0) \rangle, 
\label{eq:corr_func}
\end{equation}
%
where e.g.  from Eq. (\ref{eq:bath_shift}),
\bea
{\hat B_L}^{-\xi}(t)=
\sum_k  \lambda_{L,k}
\left[\hat b^\dagger_{L,k}(t)e^{-i\omega_{L,k}\xi/2}+\hat b_{L,k}(t)e^{i\omega_{L,k}\xi/2}\right]. 
\label{eq:Bxi}
\nonumber\\
\eea
%
We explicitly evaluate Eq. (\ref{eq:corr_func}) and get
\begin{equation}
\eta_\alpha(t+\xi)
=\frac{1}{2\pi}
\int_{-\infty}^{\infty}d\omega \frac{g_\alpha(\omega)
e^{\frac{\beta_\alpha\omega}{2}}}{\sinh\frac{\beta_\alpha\omega}{2}}e^{-i\omega(t+\xi)},
\label{eq:cont_corr}
\end{equation}
where we extend the range of the spectral density to negative frequencies, $g(\omega)=-g(-\omega)$.
We conclude that the impact of the counting field is a multiplicative phase factor onto
the standard integrand. 

To evaluate Eq. (\ref{eq:PI}) we time-discretize the influence functional 
and accompanying correlation functions. 
This is accomplished in the same way as in Ref. \cite{Makri1995}, and we receive 
\begin{widetext}
\begin{equation}
\begin{split}
I^\xi_\alpha({\{s^{\pm}\}}) &= \exp \Bigg[-\sum_{k=0}^{N} \sum_{k'=0}^{k}\, \Big(s_k^{+}s_{k'}^{+}{\eta}^{++}_{kk',\alpha}-s_k^{-}s_{k'}^{+}\eta^{-+,\xi}_{kk',\alpha}-s_k^{+}s_{k'}^{-}\eta^{+-,\xi}_{kk',\alpha}+s_k^{-}s_{k'}^{-}\eta^{--}_{kk',\alpha}  \Big)\Bigg].\\
\end{split}
\label{eq:disc_IF}
\end{equation}
\end{widetext}
The $\eta^{\pm\pm,\xi}_{kk',\alpha}$ coefficients are the discretized versions 
of Eq. (\ref{eq:cont_corr}).
The influence functional contains four different types of correlation functions, 
rather than the two that appear in the standard derivation for the reduced density matrix \cite{Makri1995},
see Appendix A.
The coefficients $\eta^{++}$ and $\eta^{--}$ do not depend on the counting field, 
and are complex conjugates, $(\eta^{++}_{kk'})^*=\eta^{--}_{kk'}$.
The two new correlation functions, $\eta^{-+,\xi}$ and $\eta^{+-,\xi}$ 
are dressed by the counting field, and
they approach $\eta^{++}$ and $\eta^{--}$, respectively, at $\xi=0$ when
one retrieves the standard influence functional for the reduced density matrix.
Note that  $\eta^{-+,\xi}\neq\big(\eta^{+-,\xi}\big)^*$,  thus
the counting field destroys the blip-sojourn symmetry of 
the standard influence functional underlying recent blip  
decomposition  based path integrals \cite{Makri2017}.

\subsection{Iterative time evolution}

Attempting to evaluate Eq. (\ref{eq:PI}) directly is impractical,
in particular, when looking for steady-state properties. 
The cost scaling of this full path integral is exponential, 
$d^{2N}$, where $d$ is the dimensionality of the Hilbert state of the 
system  and $N$ is the number of time steps.
We therefore implement an iterative scheme, in the mold of the celebrated iQuAPI approach.

Following the reasoning of Makri and Makarov \cite{Makri1995}, the correlation function (or kernel) 
$\eta(t)$ has a finite range (captured by a decorrelation timescale) if the environment is characterized by a smooth and continuous spectrum.
This observation allows the development of 
an iterative time evolution scheme that is numerically exact when 
sufficiently long memory, $\Delta k=k-k'$, 
is accounted for, and when the Trotter error is minimized, $\delta t \to 0$. 
The introduction of the counting field simply shifts the time axis. 
Therefore, the same principles underlying the development of the iQuAPI apply to the iFCSPI method,
and by propagating the path integral using the counting field dependent
influence functional we should converge to the exact generating function.

We now provide the working expressions for the iterative time evolution scheme.
First, recall the definition of counting field dependent
reduced density matrix and the characteristic function,
\bea
\sigma^{\xi}(t) \equiv  {\rm Tr_B}[ \rho^{\xi}(t)], \,\,\,\,
{\mathcal Z}(t,\xi) = {\rm Tr_S}[ \sigma^{\xi}(t)].
\eea
We define the truncated IF, which is used for time evolution,
\begin{widetext}
\bea
I^{\Delta k,\xi}_{\alpha}(s_{k-\Delta k}^{\pm},s_{k-\Delta k+1}^{\pm},...,s_{k}^{\pm}) = 
 \exp \Big[-\sum_{k'=k-\Delta k }^{k}\, \Big(s_k^{+}s_{k'}^{+}{\eta}^{++}_{kk',\alpha}-s_k^{-}s_{k'}^{+}\eta^{-+,\xi}_{kk',\alpha}-s_k^{+}s_{k'}^{-}\eta^{+-,\xi}_{kk',\alpha}+s_k^{-}s_{k'}^{-}\eta^{--}_{kk',\alpha}  \Big)\Big].
\label{eq:truncIF}
\eea
\end{widetext}
It includes the pairwise coupling terms between  $s_k$ and previous time steps within the memory time.
Time evolution is dictated by 
\bea
&&\sigma_{\Delta k}^{\xi}(s_{k-\Delta k}^{\pm},s_{k-\Delta k+1}^{\pm},... ,s_{k}^{\pm})=
\nonumber\\
&&\sum_{s_{k-\Delta k-1}^{\pm}} \sigma_{\Delta k}^{\xi}(s_{k-\Delta k -1}^{\pm},s_{k-\Delta k}^{\pm},... ,s_{k-1}^{\pm}) 
K(s_{k-1}^{\pm}, s_{k}^{\pm}) 
\nonumber\\
&&\times
I_{L}^{\Delta k,\xi/2}(s_{k-\Delta k}^{\pm},...,s_{k}^{\pm}) I_{R}^{\Delta k,-\xi/2}(s_{k-\Delta k}^{\pm},...,s_{k}^{\pm}),
\nonumber\\
\label{eq:TE}
\eea
where we introduce the augmented reduced density matrix  $\sigma^\xi_{\Delta k}$ (also known as the reduced density tensor), 
 which carries the path information within $\tau_m=\Delta k \delta t$. 
It is initialized by performing exact time evolution up to $t_{\Delta k}=\Delta k\delta t$, without any summation.
To get the time-local value, we construct an additional truncated IF as in Eq. (\ref{eq:truncIF}), 
but with the final time step $k=N\delta t$ (see Appendix A). We use this truncated IF 
in the last time evolution step Eq. (\ref{eq:TE}), and sum over the intermediate times,
\bea
\sigma^{\xi}(t_k) = \sum_{s_{k-\Delta k}^{\pm},...,s_{k-1}^{\pm}} 
\sigma_{\Delta k}^{\xi}(s^{\pm}_{k-\Delta k },... ,s_k^{\pm}).  
\eea
The counting-dressed reduced density matrix has four elements. We trace over the diagonal elements, $s_k=s_k^+=s_k^-$
to construct the characteristic function of heat exchange at time $t_k$,
$\mathcal Z(t_k,\xi)= \sum_{s_k=\pm}\langle s_k |\sigma^{\xi}(t_k)|s_k \rangle$.

Examining the form of Eq. (\ref{eq:disc_IF}), 
it is trivial to show that  
${I}^{\xi=0}_\alpha ({\{s^{\pm}\}})={I}_\alpha ({\{s^{\pm}\}})$. 
Therefore, from the trace conservation  that is inherent to the iQuAPI algorithm we conclude that
 ${\cal Z}(t,\xi=0)=1$. 

\begin{figure*}[htpb]
\makebox[\textwidth][c]{\includegraphics[scale=.5] {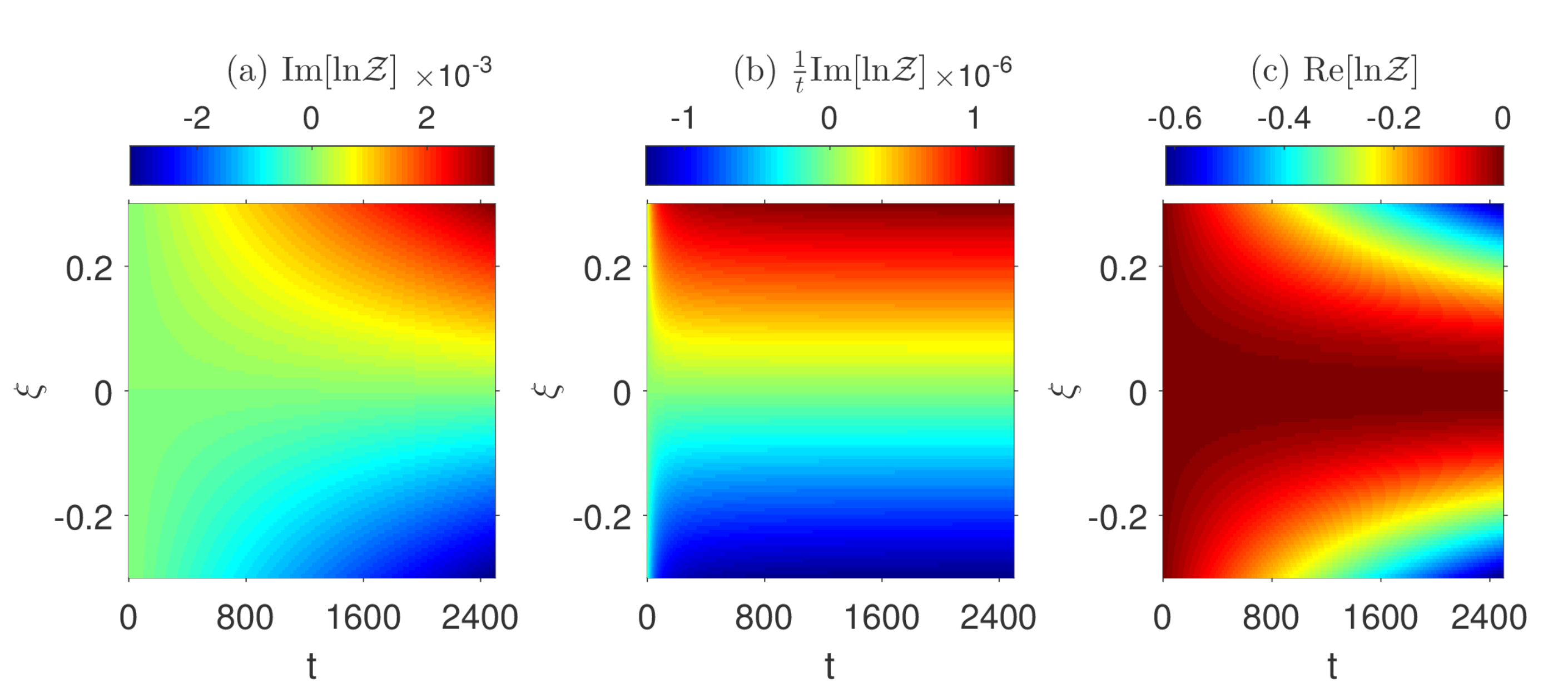}} 
\caption{Representative examples of the
imaginary (a)-(b) and real (c) components of the
generating function for heat exchange in the NESB model.
Panel (b) displays the imaginary part of the cumulant generating function
as it reaches steady state. Results in panel (c) are not fully converged,
and should be considered qualitative.
Parameters are $\Delta=1$, $\omega_c=10$, $\bar{T}=5$, $T_L-T_R=0.05$, $\gamma=10^{-3}$, $\delta t=0.3$, $\Delta k=5$.}
\label{fig:PI_characteristics}
\end{figure*}


\section{Results}
\label{Sec-results}

\subsection{Choice of parameters}
The iterative path integral method provides the exact answer when the time step is infinitely short, $\delta t \to0$, 
and the entire range of the memory kernel is covered, $\Delta k=N$.
In practice, we minimize the Trotter error by taking a small enough time step $\delta t$, 
while aiming to adequately capture the memory of the bath, 
by increasing $\Delta k$, which controls the range of the memory retained. 
Convergence is approached by adjusting these parameters until the result 
remains constant for multiple iterations. 

Balancing the Trotter error with the memory-truncation error is tricky:
Shortening the time step increases the number of steps required to span the memory kernel.
Since  the computational effort scales as $d^{2\Delta k}$, with $d$ the dimension of the Hilbert 
space of the system, simulations with the basic iQuAPI algorithm 
are typically limited to $\Delta k \lesssim 10$ \cite{Makri1995}. 
Advanced algorithms based on path filtering \cite{Lambert2012} 
or tensor decomposition \cite{Strathearn2018}, for example, extend 
this limit by an order of magnitude or more---at the cost of additional algorithmic 
complexity and introduction of new convergence parameters. 
In particular, filtering approaches are not trace conserving.
Other approaches to improving on the basic iQuAPI 
include those based on blip decomposition 
\cite{Makri2017} or mixed quantum-classical setups \cite{MakriQC1,MakriQC2}, 
but are not compatible with the iFCSPI algorithm. 
To lucidly present our method and its computational 
challenges, we therefore limit ourselves to the basic iQuAPI algorithm of Refs. \cite{Makri1995,Makri1995a}. 

To facilitate convergence, our simulations are typically performed using the following parameters:
spin splitting $\Delta=1$,  and high averaged temperature, $\bar{T}=(T_L+T_R)/2=5$.
We assume an Ohmic spectral density function
for the baths and work in the scaling limit with a high cutoff frequency $\omega_c=10-50$.
Since the iFCSPI method is not restricted to the linear response regime 
(and in fact, converges well far from equilibrium),
we play with biases in the range $T_L-T_R=0-0.5$.
For the dimensionless system-bath coupling we use $\gamma=10^{-3}-1$, but convergence was 
achieved only at weak-to-intermediate coupling, $\gamma\lesssim 0.15$.  
With these parameters, iFCSPI simulations typically converge
with $\Delta k=4-7, \delta t=0.05-0.2$. 
We also define the memory span $\tau_m\equiv\Delta k \delta t$. Roughly, we find 
that iFCSPI simulations converge when  $\tau_m\gtrsim1/\bar T$. 

The iFCSPI protocol has comparable convergence characteristics to iQuAPI.
We found that it is slightly more difficult to converge the
imaginary part of the CGF than the reduced density matrix,
and much more difficult to converge the real part of the CGF. After presenting as an example 
the CGF, in the rest of the paper we focus on the behavior of the first cumulant of heat exchange, 
the heat current.

\subsection{Perturbative methods}
In order to assess the  viability of the iFCSPI
we compared its results to several other methods, which are perturbative in the interaction parameter
$\alpha$ or the tunneling element $\Delta$:

(i) The Markovian Redfield quantum master equation (QME) is a second order perturbative approach, 
valid for a system weakly coupled to a Markovian environment \cite{Nitzan2006}.
The resulting heat current expression only includes resonant heat exchange processes. 
We use expressions developed in Refs. \cite{segal-PRL,segal-QME}, 
and more recently reviewed in Ref. \cite{Agarwalla2017a}.
We refer to this approach as the ``Redfield QME'' method.

(ii) The Majorana fermion Green's function (Majorana-GF) approach is
perturbative is the system-bath coupling, 
but it can capture non-resonant processes beyond the Redfield QME. 
The working expressions were organized in Ref. \cite{Agarwalla2017a}. 

(iii) Polaron transforming the spin-boson model, one can derive expressions for the heat current that
are related to the
noninteracting blip approximation (NIBA) literature \cite{Ren,Ren2,SaitoNJP}. Under the Markov approximation
we organize the NIBA QME \cite{segal-PRL,segal-QME,Nicolin2011b,Nicolin2011},
which is valid for intermediate-strong system-bath coupling when $\bar{T}\gg\Delta$.
We note that a more accurate polaron transformed method was developed more recently:
It interpolate between the Redfield limit and NIBA \cite{Ren,Ren2},
or go beyond the Markovian limit \cite{SaitoNJP}. Here we limit ourselves to 
the high temperature Markov equation that was 
described in Refs. \cite{segal-PRL,Nicolin2011}, referred below to as ``NIBA" expression for heat exchange.

\subsection{Cumulant Generating Function}

The behavior of  the function $\ln \, \mathcal{Z}(t,\xi) $ is displayed in 
Fig. \ref{fig:PI_characteristics}.
The imaginary part of the function is odd in the counting parameter.
In panel (b) we show the cumulant generating function as it approaches and reaches its 
nonequilibrium steady-state. Taking the derivative with respect to $i\xi$ provides the  averaged heat current.
The real part of $\ln \mathcal{Z}(t,\xi)$ is displayed in panel (c).
As expected, it is even in the counting parameter, 
see Eq. (\ref{eq:CGF}).
While our results for the imaginary part of $\mathcal{Z}(t,\xi)$ easily converge with respect to the time step and memory size (here with $\delta t=0.3$, $\Delta k=5$), the real part of this function requires stringent conditions for convergence, beyond what our implementation can currently meet.

We compute the steady-state current via numerical derivatives of the cumulant generating function, following Eq. (\ref{eq:J}).
The cumulant generating function reaches its steady-state limit when $\ln[\mathcal{Z}(t,\xi)]$ becomes linear in time. At that point, one evaluates a numerical derivative with respect to $t$ to extract $\mathcal{G}(\xi)$. Here $\xi$ is always the smallest timescale of the system, and particularly, small compared to $\delta t$. It is further not difficult to check the convergence with respect to $\xi$, to ensure one captures only the linear behavior of the cumulant generating function. In our simulations we typically take $\xi$ on the order $\delta t/100$.

\subsection{Heat current}

\begin{figure*}[htpb]
\makebox[\textwidth][c]{\includegraphics[scale=.5]{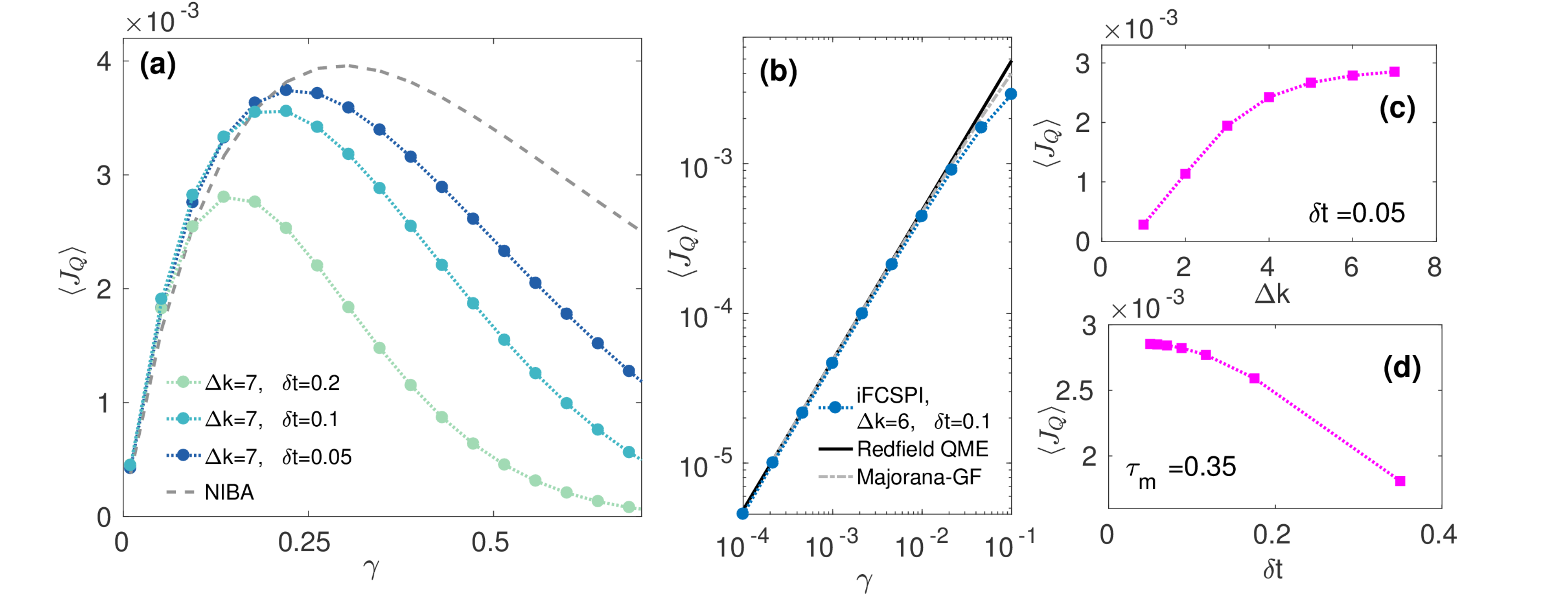}} 
\caption{ (a) Steady-state heat current in the NESB model as a function of the dimensionless 
system-bath coupling parameter $\gamma$ with $\Delta k=7$ and different time steps $\delta t$.
Results are compared to NIBA simulations.
(b) weak coupling results are compared to the Redfield QME and the Majorana Green's function approaches. 
(c)-(d) Exemplifying convergence of results at $\gamma=0.1$ 
as a function of (c)  $\Delta k$ while fixing the time step and (d) $\delta t$
for a fixed memory time.
Parameters are $\Delta=1$, $\omega_c=50$, $\bar{T}=5$, $T_L-T_R=0.5$.} 
\label{fig:jvsg}
\end{figure*}

\begin{figure*}[htpb]
\includegraphics[scale=.5]{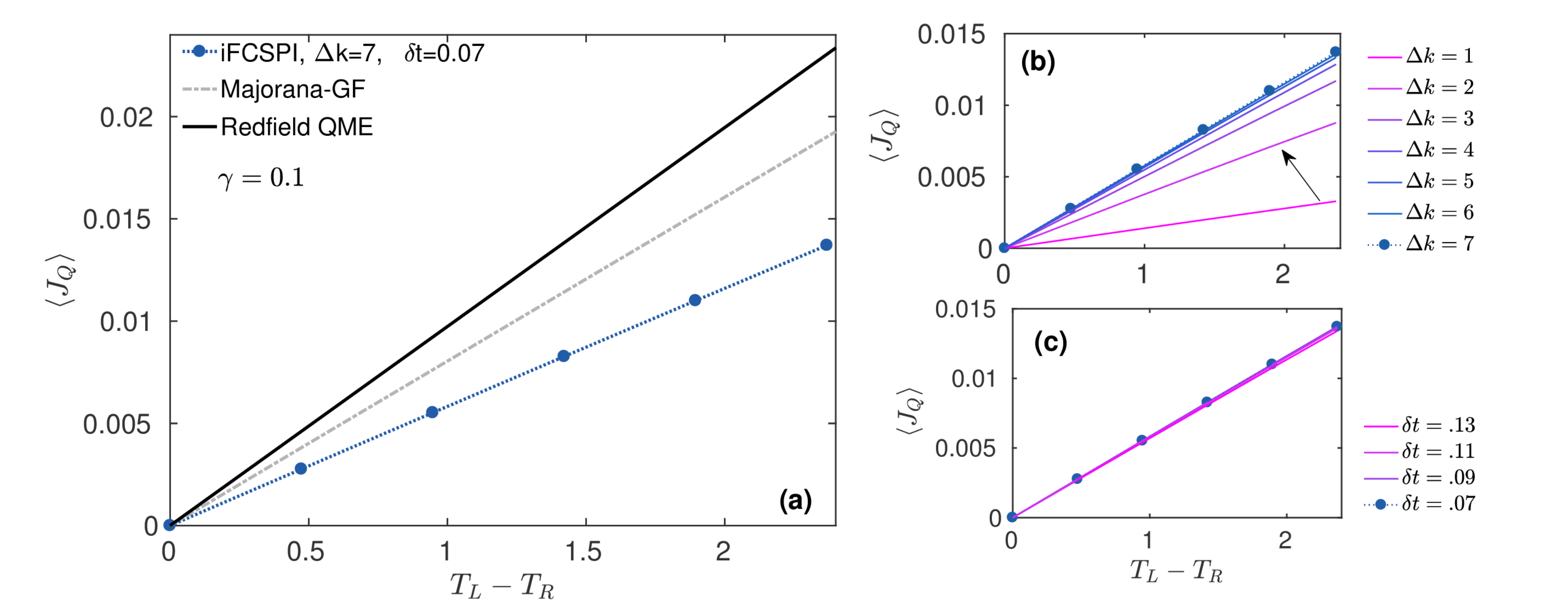} 
\caption{(a) Steady state heat current in the NESB model as a function of temperature difference 
$T_L-T_R$.
iFCSPI simulations are compared to Redfield and Majorana-GF approaches.
(b)-(c) Convergence is achieved for both small and large temperature bias as
illustrated upon varying (b) the  memory range  $\Delta k$ with a fixed $\delta t=0.07$,
 and the (c) time step $\delta t$ with a fixed $\Delta k=7$. 
The arrow in panel (b) points to the direction of converging results.
Parameters are $\Delta=1$, $\gamma=0.1$, $\omega_c=50$, $\bar{T}=5$.}
\label{fig:jvsDT}
\end{figure*}


{\it System-bath coupling.}
Figure \ref{fig:jvsg} displays the main result of this work, that is the behavior of the heat current
as a function of the coupling strength to the baths.
We compare the behavior of the iFCSPI to other approaches, and to facilitate comparison we define a dimensionless  parameter, 
$\frac{\gamma\Delta}{\bar{T}}$. We expect iFCSPI simulations to comfortably converge when $\frac{\gamma\Delta}{\bar{T}}\ll1$.
 Here, $\gamma \Delta$ is the system-bath interaction energy in the Redfield limit; the bath-induced
transition rate in the Redfield QME in proportional to this factor through its dependence on $g(\omega)$, assuming an Ohmic form.
The Redfield formalism holds for the NESB so long as the interaction
energy is assumed smaller than the energy spacing in the system, i.e. $\gamma\Delta \ll \Delta$ or $\gamma\ll1$. NIBA is valid at high temperatures
 in the nonadiabatic limit,  $\Delta <\bar T$. 

The agreement with weak-coupling approaches is excellent when $\gamma$ is small, as we observe in panel (b), while
at weak-intermediate coupling, $\gamma\gtrsim 0.1$ (or equivalently, $\frac{\gamma\Delta}{\bar{T}}\gtrsim 0.02$), 
the results deviate by about a factor of 1.5-2
as we pass out of the range of validity for the Redfield and Majorana approaches. 

As we push the iFCSPI method into the stronger system-bath coupling regime, $\gamma>0.25$ (or equivalently,
 $\frac{\gamma\Delta}{\bar{T}}> 0.05$), convergence becomes difficult
in the current implementation.
Results are compared to the Markovian NIBA  approach \cite{Nicolin2011}, 
which is known to be qualitatively correct at strong coupling for Ohmic dissipation, 
though in general it overestimates the current. 
We understand {\it a priori} that we will not be able to converge the iFCSPI at strong coupling with 
our current implementation, simply due to the prohibitive computational cost of additional system-bath memory. 
That being said, the method converges quite well up to $\gamma\simeq 0.2$, and the qualitative behavior 
of iFCSPI simulations is fair over the full range. 
The peak position is close to that from NIBA, and it follows the expected turnover behavior. 

Panels (c) and (d) in Fig. \ref{fig:jvsg} exemplify the convergence of the iFCSPI at weak-intermediate coupling. 
We first show the convergence as a function of  the memory time covered by increasing 
$\Delta k$ with a short time step. 
We then demonstrate a fixed-memory analysis, where the total memory time is kept fixed, 
$\tau_m=\Delta k\cdot \delta t$, and the number of time steps required to cover this memory range is increased, 
eventually converging with a short enough time step.


{\it Temperature difference.}
In Fig. \ref{fig:jvsDT} we study the behavior of the heat current as a function of the temperature difference,
observing a linear dependence. 
This trend agrees with weak coupling approaches across the full range of temperature biases examined;
for a symmetric junction, the Redfield approaches predicts  $\langle J_Q \rangle \propto \frac{(T_L-T_R)}{\bar T}$ \cite{segal-PRL,segal-QME,claire}.
While weak coupling methods are qualitatively correct for the presented parameters,
the Redfield QME overestimates  the heat current by almost a factor of two throughout, which is quite significant.
We note that our results fully converge both at small, $(T_L-T_R)/\Delta <1$, and large, $(T_L-T_R)/\Delta >1$,
temperature differences, see panels (b)-(c).
With algorithmic improvements, one could test the behavior of the heat current as a 
function of temperature difference at strong coupling. Furthermore, by introducing a spatial asymmetry with different
coupling energies at the two boundaries one could assess the extent of the thermal diode effect, which so far was 
predicted and evaluated based on perturbative methods \cite{segal-PRL, segal-QME}.

\begin{figure*}[htbp]
\includegraphics[scale=.5] {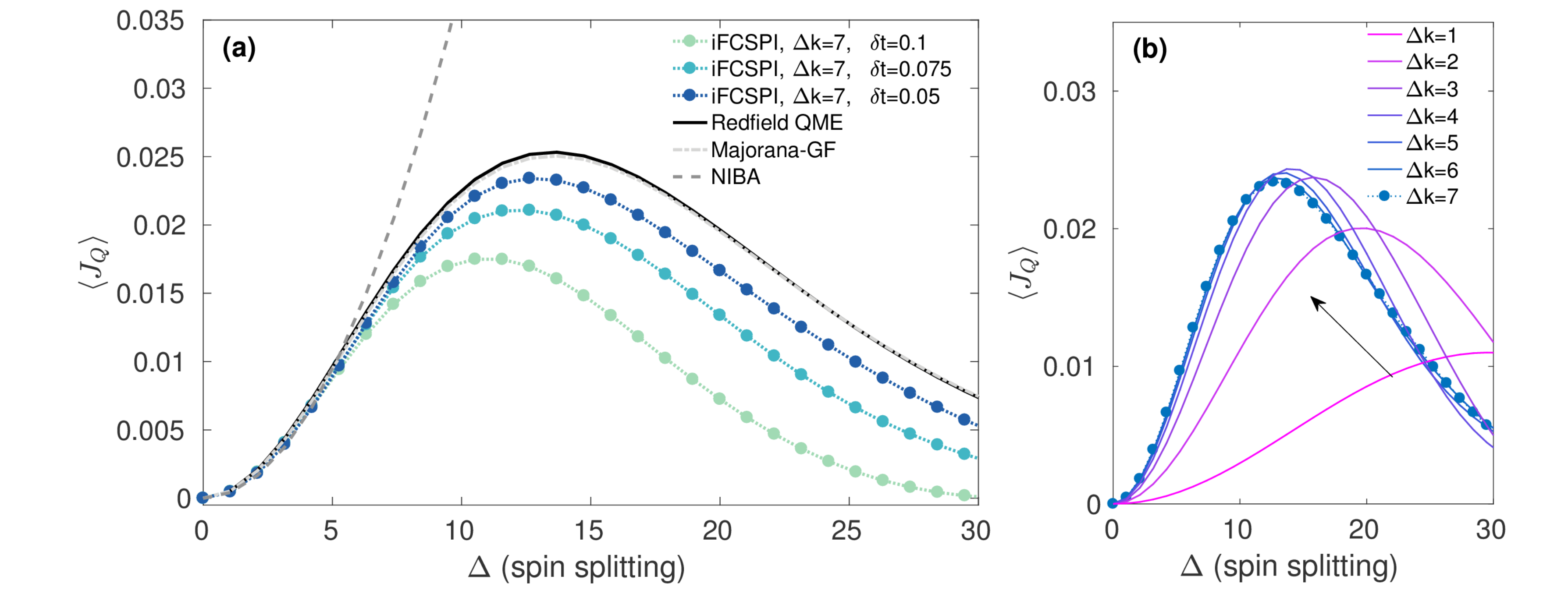} 
\caption{(a) Steady state heat current as a function of the tunneling element $\Delta$ 
in the weak system-bath coupling regime. 
We present three curves for the iFCSPI with different time steps $\delta t$.
iFCSPI results are compared to Redfield QME (full), Majorana-GF (dashed-dotted), 
and NIBA (dashed) results. The latter is known to fail in the large $\Delta$ regime. 
(b) Demonstration of convergence behavior: series of results from $\Delta k=1-6$ for $\delta t=0.05$.
The arrow points to the direction of increasing $\Delta k$. The case $\Delta k=7$, which is presented in panel (a) is highlighted in 
filled circles.
Parameters are $\gamma=0.01$, $\omega_c=50$, $\bar{T}=5$, $T_L-T_R=0.5$.}
\label{fig:jvsDelta}
\end{figure*}

{\it Coherent tunneling/level splitting.}
Fig. \ref{fig:jvsDelta} shows the behavior of the current as we increase the eigenenergy splitting, $\Delta$,
which corresponds to the tunneling energy in the site basis.
Note that we enter an effective low temperature regime as we increase $\Delta$, and it is more difficult to converge the iFCSPI when $\Delta\gtrsim \bar T$.
That being said, the iFCSPI maintains excellent agreement with  weak coupling approaches up to 
$\frac{\gamma\Delta}{\bar T} \simeq 0.02$, and is in qualitative agreement for the full range. 
The three lines shown for the path integral are each converged in terms of memory length 
($\Delta k\cdot \delta t$), and we approach the correct answer by 
successively reducing the time step $\delta t$. NIBA follows the expected $\langle J_Q\rangle \propto \Delta^2$
trend \cite{Nicolin2011}, and it breaks down when $\frac{\gamma\Delta}{\bar T}\gtrsim 0.01$.
In panel (b) we display convergence in $\Delta k$ for the shortest time step, 
and we find that the method converges for the full range of $\Delta$, for the given $\tau_m$.

\section{Conclusions}
\label{Sec-Summ}

We introduced an iterative influence functional path integral approach to simulate the full 
counting statistics of heat exchange in quantum impurity models.
This iFCSPI generalizes the celebrated iQuAPI, going beyond iterative evolution of the reduced density matrix 
to study energy transfer statistics in a numerically exact manner.

The numerical implementation of the iFCSPI relies on a discretized representation of the system's coordinates.
To test the this new approach, we applied it to the nonequilibrium spin-boson model 
and focused on the average heat current in steady state.
We studied the heat transfer behavior as a function of system-bath coupling
and tunneling splitting, demonstrating turnover behavior in both cases. The iFCSPI method
can be employed far from equilibrium at large temperature difference 
since its working expressions are not based in a linear response assumption. 
Though not elaborated on in this work, one can employ the iFCSPI to retrieve transient energy exchange statistics by considering the behavior of the generating function between its initial condition and the steady-state limit.

Key challenges associated in converging iFCSPI simulations were elaborated on. Most critically, 
Trotter splitting leads to spurious heat current at zero temperature difference when 
the time step $\delta t$ is finite. Nevertheless,
this error is precisely canceled out in junctions with identical bath spectral functions and upon symmetrizing the counting fields at the two boundaries.
Besides numerical exactness, a key benefit of the iFCSPI approach is the ease of implementation,
coming as a straightforward modification to the iQuAPI method.
Converging the path integral for parameters in the range
$\omega_c>\bar T>\Delta>\gamma\Delta$ was feasible even with the basic iQuAPI algorithm.

Beyond the spin-boson model, the principles outlined here can be used to study other
models for charge and energy transport based on a path integral representation of the 
generating function, constructed similarly to Eq. (\ref{eq:PI}).
For example, FCS of the interacting Anderson dot model \cite{Thorwart1,Thorwart2,Guy18,Stegmann1,Stegmann2} 
or the spin-fermion model can be similarly implemented using an iterative algorithm.

Future work will be focused on implementing the iFCSPI within efficient path integral algorithms developed in the area of dissipative dynamics, 
such as path filtering \cite{Lambert2012} and tensor decomposition \cite{Strathearn2018}. 
Coupled with a sufficiently powerful algorithm, the iFCSPI is a tremendously enabling tool, handing over not only the system reduced density matrix, but also the energy exchange statistics to numerically exact precision. 
An improved iFCSPI algorithm would enable the exploration of non-Markovianity, strong system-bath coupling effects, quantum coherences and time-dependent driving, offering itself as a powerful instrument for studies of quantum thermal machines.
%

\begin{acknowledgments}
D.S. acknowledges the Natural Sciences and Engineering Research Council (NSERC) of Canada Discovery Grant
and the Canada Research Chairs Program.
The work of M.K. was supported by an NSERC Canada Graduate Scholarship-Doctoral (CGS D).
B. K. A. acknowledges support from the Centre for Quantum Information and Quantum Control 
and the hospitality of the Department of Chemistry at the  University of Toronto.
\end{acknowledgments}


\begin{widetext}
\renewcommand{\theequation}{A\arabic{equation}}
\setcounter{equation}{0}  
\section*{Appendix A: Counting-field dressed influence functionals for general baths}

The influence functional (\ref{eq:IF}) was derived in Ref. \cite{Weiss} following the original derivation
of the Feynman-Vernon influence functional.
The derivation relies on harmonic baths and a bilinear interaction Hamiltonian 
in the displacements of the reservoirs.
Here, we construct the counting field-dressed influence functional 
using perturbation theory and re-exponentiation. 
This procedure does not rely on the details of the model, and it therefore applies to more general cases,
e.g. anharmonic baths.

\subsection{IF for a single bath}
For simplicity, we begin by describing the IF assuming the system is coupled to a single bath, $L$.
The influence functional is derived by focusing on the time-dependent environmental Hamiltonian
$\hat H_{env}=\hat H_L + \hat H_{SL}(t)$, which evolves along the system's path $s(t)$ as
 $\hat{H}_{SL}(t)=s(t)\hat B$.
We later generalize our results to the case of two baths.

We begin with the formal expression for the total influence functional,
\begin{equation}
\begin{split}
I^\xi_L(\{s^\pm\}) 
={\rm Tr}_B\left[\rho_B\hat{U}^{\dagger,\xi}_{env}[s^-(t)]\hat{U}^{-\xi}_{env}[{s^+(t)}]\right],
\end{split}
\label{eq:App1}
\end{equation}
where e.g. $\hat{U}^{-\xi}_{env}[s^+(t)]={\rm T} e^{-i\int_0^t\hat H_{env}^{-\xi}(s^+,\tau) d\tau}$
is the interaction picture propagator of the environmental Hamiltonian moving on the forward $(+)$  path.
Here, we define the interaction picture operators as
%
\bea
\hat{H}_{SL}^{-\xi}(s^+,t)\equiv s^+(t)\hat B_L^{-\xi}(t), \,\,\,
\hat{H}^{-\xi}_{env}(s^+,t) = \hat{H}_L+\hat{H}_{SL}^{-\xi}(s^+, t).
\eea
In this equation, the time dependence $\hat B(t)$ results from the interaction picture, while $s(t)$ corresponds to the time dependent coordinate of the system (path) along which the bath coordinates evolve.
To simplify our notation, we omit the time argument of
the system coordinate, $s^{\pm}(t) \to s^{\pm}$ until we arrive at Eq. (\ref{eq:A6}), where we recover it.
Similar definitions hold for the backward time evolution along the system coordinate $s^-$, preparing the 
time-dependent environmental Hamiltonian $H_{env}^{\xi}(s^-,t)$. 

Expression (\ref{eq:App1}) can be evaluated by a Dyson expansion and a re-exponentiation procedure as we show next, or by explicit consideration of a time-dependent force on a harmonic oscillator, as in Ref. \cite{Schwinger1961,Weiss}. 
%
We now explicitly write the time evolution operators as a Dyson series,
%
\begin{equation}
\begin{split}
\hat{U}^{-\xi}_{env}(s^+,t) &= e^{-i\hat{H}_Lt} \Bigg[1-i\int_0^td\tau\hat{H}^{-\xi}_{SL}(s^+,\tau) - \int_0^t\int_0^{\tau}d\tau d\tau'\hat{H}^{-\xi}_{SL}(s^+,\tau)\hat{H}^{-\xi}_{SL}(s^+,\tau')+...\Bigg]  \\
\hat{U}^{\dagger,\xi}_{env}(s^-,t) &= \Bigg[1+i\int_0^td\tau\hat{H}^\xi_{SL}(s^-,\tau) - \int_0^t\int_0^{\tau}d\tau d\tau'\hat{H}^\xi_{SL}(s^-,\tau')\hat{H}^\xi_{SL}(s^-,\tau)+...\Bigg]e^{i\hat{H}_Lt}.
\end{split}
\end{equation}
Multiplying the two propagators and keeping terms up to second order, then rearranging, we retrieve
\bea
\hat{U}^{\dagger,\xi}_{env}(s^-,t)\hat{U}^{-\xi}_{env}(s^+,t)&=&
1-i\int_0^td\tau\left[\hat{H}^{-\xi}_{SL}(s^+,\tau)-\hat{H}^{\xi}_{SL}(s^-,\tau)\right]
+\int_0^td\tau\hat{H}^{\xi}_{SL}(s^-,\tau)\int_0^td\tau\hat{H}^{-\xi}_{SL}(s^+,\tau)
\nonumber\\
&-&\int_0^t\int_0^\tau d\tau d\tau'\left[\hat{H}^{\xi}_{SL}(s^-,\tau')\hat{H}^{\xi}_{SL}(s^-,\tau)+\hat{H}^{-\xi}_{SL}(s^+,\tau)\hat{H}^{-\xi}_{SL}(s^+,\tau') \right] + ...
	\eea
The third term (product of two integrals) can be rearranged, resulting in
\bea
\hat{U}^{\dagger,\xi}_{env}(s^-,t)\hat{U}^{-\xi}_{env}(s^+,t)
&=&1-i\int_0^td\tau\left[\hat{H}^{-\xi}_{SL}(s^+,\tau)-\hat{H}^{\xi}_{SL}(s^-,\tau)\right]
\nonumber\\
&+&\int_0^t\int_0^\tau d\tau d\tau' \left[\hat{H}^{\xi}_{SL}(s^-,\tau)\hat{H}^{-\xi}_{SL}(s^+,\tau')+\hat{H}^{\xi}_{SL}(s^-,\tau')\hat{H}^{-\xi}_{SL}(s^+,\tau)\right]
\nonumber\\
&-&\int_0^t\int_0^\tau d\tau d\tau'\left[\hat{H}^{\xi}_{SL}(s^-,\tau')\hat{H}^{\xi}_{SL}(s^-,\tau)+\hat{H}^{-\xi}_{SL}(s^+,\tau)\hat{H}^{-\xi}_{SL}(s^+,\tau') \right]+ ...
\eea
Next, we evaluate the trace over the bath with averages performed over the thermal state $\rho_B$, and assuming that the 
thermal average of the interaction vanishes, $\langle \hat H_{SL} \rangle=0$. Relying on the bilinear form of the interaction, e.g.
$\hat H_{SL}^{-\xi}(s^+,t)=s^+(t)\hat B_{L}^{-\xi}(t)$,  we get
\bea
{I}^\xi_L(\{s^\pm\})&=&
1+\int_0^t\int_0^\tau d\tau d\tau'
\Big[
s^-(\tau)s^+(\tau')\langle\hat{B}^{\xi}_{L}(\tau)\hat{B}^{-\xi}_{L}(\tau')\rangle
+s^+(\tau)s^-(\tau')\langle\hat{B}^{\xi}_{L}(\tau')\hat{B}^{-\xi}_{L}(\tau)\rangle
\nonumber\\
&-&s^+(\tau)s^+(\tau')\langle\hat{B}^{-\xi}_{L}(\tau)\hat{B}^{-\xi}_{L}(\tau')\rangle
-s^-(\tau)s^-(\tau')\langle\hat{B}^{\xi}_{L}(\tau')\hat{B}^{\xi}_{L}(\tau)\rangle\Big] + ...
\label{eq:A6}
\eea
%
Recall that bath operators are written here in the interaction representation. The time-dependence of
the system's coordinate corresponds to the system's path on which the bath evolves.

Using the stationary bath condition and the cyclic property of the trace
we lose the counting parameter in correlation functions of the form 
\bea
\langle\hat{B}^{\xi}_{L}(\tau')\hat{B}^{\xi}_{L}(\tau)\rangle&=&
{\rm Tr}_B
\big[e^{i\hat{H}_L\xi/2}\hat{B}_{L}(\tau')e^{-i\hat{H}_L\xi/2}
e^{i\hat{H}_L\xi/2}\hat{B}_{L}(\tau)e^{-i\hat{H}_L\xi/2}\rho_B \big]
\nonumber\\
&=&{\rm Tr}_B \big[
\hat{B}_{L}(\tau') \hat{B}_{L}(\tau)\rho_B \big].
\eea
In contrast, the counting parameter survives in the following combination,
\bea	
\langle\hat{B}^{\xi}_{L}(\tau')\hat{B}^{-\xi}_{L}(\tau)\rangle&=&
{\rm Tr}_B\big[e^{i\hat{H}_L\xi/2}\hat{B}_{L}(\tau')e^{-i\hat{H}_L\xi/2}
e^{-i\hat{H}_L\xi/2}\hat{B}_{L}(\tau)e^{i\hat{H}_L\xi/2}\rho_B \big]
\nonumber\\
&=&{\rm Tr}_B\big[e^{i\hat{H}_L\xi}\hat{B}_{L}(\tau'-\tau)e^{-i\hat{H}_L\xi}\hat{B}_{L}(0)\rho_B \big]
\nonumber\\
&=&\langle \hat{B}_{L}(\tau'-\tau+\xi)\hat{B}_{L}(0)\rangle.
\eea
So far, the derivation only includes terms up to second order in the system-bath interaction. 
Fortunately, for a bilinearly-position coupled harmonic bath, the re-exponentiation of this expression is exact. 
Therefore, the continuous-time counting-dressed influence functional is given as
\begin{equation}
\begin{split}
I_L^\xi(\{s^\pm\})
=\exp\Big[-\int_0^td\tau\int_0^\tau d\tau'\big[&s^+(\tau)s^+(\tau')\eta_L(\tau-\tau')-s^-(\tau)s^+(\tau')\eta_L(\tau-\tau'+\xi)\\
-&s^+(\tau)s^-(\tau')\eta_L(\tau'-\tau+\xi)+s^-(\tau)s^-(\tau')\eta_L(\tau'-\tau)\big]\Big].
\end{split}
\end{equation}
Using the interaction Hamiltonian (\ref{eq:Bxi}), we get the correlation function,
\bea
\eta_L(\tau-\tau'+\xi)&=&\langle\hat{B}_{L}(\tau-\tau'+\xi)\hat{B}_{L}(0)\rangle
\nonumber\\
&=&
 \Big\langle \sum_k \lambda_{L,k}^2 
\left(  \hat b_{L,k}^{\dagger} e^{i\omega_{k,L}(\tau-\tau'+\xi)}  + \hat b_{L,k}e^{-i\omega_{k,L}(\tau-\tau'+\xi)} \right)
\left( \hat b_{L,k}^{\dagger} + \hat b_{L,k} \right) \Big\rangle
\nonumber\\
&=&
\sum_k \lambda_{L,k}^2 \left \{n_{L}(\omega_{L,k}) e^{i\omega_{k,L}(\tau-\tau'+\xi)} + 
[n_{L}(\omega_{L,k}) +1]e^{-i\omega_{k,L}(\tau-\tau'+\xi)}\right\}
\nonumber\\
&=&
\frac{1}{\pi}\int_0^{\infty} d\omega g_L(\omega) \left\{ [2n_{L}(\omega)+1] \cos \omega(\tau-\tau'+\xi) 
-i\sin \omega(\tau-\tau'+\xi)\right\}
\nonumber\\
&=&
\frac{1}{\pi}\int_0^{\infty} d\omega g_L(\omega) 
\frac{e^{\beta_L\omega/2}}{ e^{\beta_L\omega/2} - e^{-\beta_L\omega/2}} e^{ -i \omega(\tau-\tau'+\xi)}
%
+\frac{1}{\pi}\int_0^{\infty} d\omega g_L(\omega) \frac{e^{-\beta_L\omega/2}}{ e^{\beta_L\omega/2} - e^{-\beta_L\omega/2}} 
e^{ i \omega(\tau-\tau'+\xi)}
\nonumber\\
&=&
\frac{1}{2\pi}\int_{-\infty}^{\infty}d\omega 
\frac{g_L(\omega)e^{\frac{\beta_L\omega}{2}}}{\sinh\left(\frac{\beta_L\omega}{2}\right)}e^{-i\omega(\tau-\tau'+\xi)},
\eea
where $n_L(\omega_{L,k})=[e^{\beta_L\omega_{L,k}}-1]^{-1}$ is the Bose Einstein distribution function, and   we extended the spectral density function to the negative domain, $g_L(\omega)=-g_L(-\omega)$.
We now discretize the path, e.g. in the forward direction,
\bea s^+(t)=s_0^+\Theta(t-t_0) + \sum_{k=1}^{N} (s_k^+-s_{k-1}^+)\Theta(t-t_k),\eea
with the Heaviside function $\Theta(t)$ and the time series $t_0=0$, $t_1=(1-1/2)\delta t$, $t_2=(2-1/2)\delta t$,... $t_{N-1}=(N-1/2)\delta t$, $t_N=N\delta t$, and obtain the discrete IF,
%
\begin{equation}
\begin{split}
I^\xi_L({\{s^{\pm}\}}) &= 
\exp \left[-\sum_{k=0}^{N} \sum_{k'=0}^{k}\, \left(s_k^{+}s_{k'}^{+}{\eta}^{++}_{kk',L}-s_k^{-}s_{k'}^{+}\eta^{-+,\xi}_{kk',L}-s_k^{+}s_{k'}^{-}\eta^{+-,\xi}_{kk',L}+s_k^{-}s_{k'}^{-}\eta^{--}_{kk',L}  \right)\right].\\
\end{split}
\label{eq:Adisc_IF}
\end{equation}
The coefficients are given by
%
\begin{equation}
\begin{split}
\eta^{\mu\nu,\xi}_{kk',L}
&=\frac{2}{\pi}\int_{-\infty}^{\infty}d\omega A_L^{\mu\nu,\xi}(\omega)\sin^2(\omega\delta t/2)e^{-i\omega\delta t(k-k')}, 
\,\,\, 0<k'<k<N \\
\eta^{\mu\nu,\xi}_{kk,L}
&=\frac{1}{2\pi}\int_{-\infty}^{\infty}d\omega A_L^{\mu\nu,\xi}(\omega)(1-e^{-i\omega\delta t}), \,\,\, 0<k<N, \\
\eta^{\mu\nu,\xi}_{N0,L}
&=\frac{2}{\pi}\int_{-\infty}^{\infty}d\omega A_L^{\mu\nu,\xi}(\omega)\sin^2(\omega\delta t/4)e^{-i\omega(N\delta t-\delta t/2)}, \\
\eta^{\mu\nu,\xi}_{00,L}
&=\eta^{\mu\nu,\xi}_{NN}=\frac{1}{2\pi}\int_{-\infty}^{\infty}d\omega A_L^{\mu\nu,\xi}(\omega) \left(1-e^{-i\omega\delta t/2}\right), \\
\eta^{\mu\nu,\xi}_{k0,L}
&=\frac{2}{\pi}\int_{-\infty}^{\infty}d\omega A_L^{\mu\nu,\xi}(\omega)
\sin(\omega\delta t/4)\sin(\omega\delta t/2)e^{-i\omega(k\delta t-\delta t/4)}, \,\,\,  0<k<N \\
\eta^{\mu\nu,\xi}_{Nk',L}
&=\frac{2}{\pi}\int_{-\infty}^{\infty}d\omega A_{L}^{\mu\nu,\xi}(\omega) \sin(\omega\delta t/4)\sin(\omega\delta t/2)e^{-i\omega(N\delta t-k'\delta t-\delta t/4)}, \,\,\, 0<k'<N.
\end{split}
\label{eq:corr_funcs}
\end{equation}
Here $\mu=\pm1 $,  and $\nu=\pm1$ correspond to the different (forward and backward) combinations.
We put together the spectral function of the reservoir, the thermal factors, and the counting-field phase elements
into the function
\begin{equation}
A_L^{\mu \nu,\xi}(\omega)=\frac{g_L(\omega)e^{\frac{\nu\beta_L\omega}{2}}}{\omega^2\sinh\left(\frac{\beta_L\omega}{2}\right)}e^{-i\omega\xi(\nu-\mu)/2},
\,\,\, \mu=\pm1, \,\,\, \nu=\pm1.
\label{eq:AppA}
\end{equation}


It is important to note that 
$\eta(\tau-\tau'+\xi)\neq[\eta(\tau'-\tau+\xi)]^*$,
 equivalent to $\eta^{-+,\xi}\neq\eta^{+-,\xi}$, 
therefore one must compute these different combinations explicitly.
This further precludes a blip decomposition, since the influence functional does not neatly reorganize into a blip-sojourn form:
the influence functional with the counting field couples all time points to all others, 
regardless of their blip or sojourn character. 
Next, we generalize this result in two ways.

\subsection{Generalization I: Two baths with asymmetric counting}

Here, we couple the system to two heat baths, but we count energy only in one of the terminals, say $L$.
The environmental Hamiltonian for the forward time evolution branch is given by
\bea
\hat{H}^{-\xi}_{env}(s^+,t) &= \hat{H}_L
+\hat{H}_R+ \hat{H}_{SL}^{-\xi}(s^+,t)+\hat{H}_{SR}(s^+,t). 
\eea
An analogous expression holds for the backward time evolution expression.
The influence functional is  given by
\begin{equation}
\begin{split}
I^\xi(\{s^\pm\}) 
={\rm Tr}_B\left[\rho_B\hat{U}^{\dagger,\xi}_{env}[s^-(t)]\hat{U}^{-\xi}_{env}[s^+(t)]\right], 
\end{split}
\label{eq:App2}
\end{equation}
yet since the initial condition of the total bath is factorized,  $\rho_B=\rho_L \otimes \rho_R$,
the IF is given by a product of two independent terms,
\bea
I^\xi(\{s^\pm\}) = I_L^\xi(\{s^\pm\}) I_R(\{s^\pm\}).
\eea
The IF without counting, $I_R(\{s^\pm\})$, corresponds to the IF [Eq. (\ref{eq:Adisc_IF})] evaluated at  $\xi=0$.

\subsection{Generalization II: Two baths with symmetric counting}

As we explain in the main text and in Appendix C, we symmetrize the definition of the moment generating function.
Therefore, we count energy transfer at both the $L$ and $R$ terminals, and
the environmental Hamiltonian is given by
\bea
\hat{H}^{-\xi}_{env}(s^+,t) &= 
\hat{H}_L +\hat{H}_R+ \hat{H}_{SL}^{-\xi/2}(s^+,t)+\hat{H}_{SR}^{\xi/2}(s^+,t)
\nonumber\\
\eea
The derivation outlined previously for the IF of the $L$ bath can be repeated for the $R$ bath, 
only with the counting parameter halved and its sign switched throughout all the expressions (\ref{eq:App1})-(\ref{eq:AppA}).
Again, based on the factorized initial condition the total influence function is separable
\begin{equation}
I^\xi(\{s^\pm\})= I_L^{\xi/2}(\{s^\pm\}) I_R^{-\xi/2}(\{s^\pm\}),
\label{eq:App3}
\end{equation}
where each term follows Eqs. (\ref{eq:Adisc_IF})-(\ref{eq:corr_funcs}). 
\end{widetext}


\renewcommand{\theequation}{B\arabic{equation}}
\setcounter{equation}{0}  
\section*{Appendix B: Counting field dressed Feynman Vernon influence functional }

The characteristic function in a  path integral form was previously derived by Carrega et al. \cite{Weiss} as part of a primarily analytical work on time-dependent heat exchange in the spin-boson model. Their work generalizes the standard-dissipative Feynman-Vernon influence functional to count heat exchange processes in system-boson bath models. That derivation relies on the Gaussian property of harmonic oscillators, which allows
tracing out exactly the degrees of freedom of the bath.
Though the authors in \cite{Weiss} considered both a different picture of the two-time measurement protocol and an alternate derivation of the influence functional from Appendix A, we show here that we arrive at the same result in the continuous time limit. For simplicity, we limit this discussion to include a single heat bath, though the result is general.

First, following  \cite{Weiss} we break the influence functional into so-called ``normal",  $I^{(0)}$, and ``counting'',  $I^{(c)}$, parts,
\begin{equation}
I^\xi=I^{(0)} I^{(c)}.
\end{equation}
The normal part is the original influence functional, without any effects from the counting field, or simply the influence functional one retrieves when $\xi=0$ at which point the counting part is unity.
The normal part $I^{(0)}$ in Ref. \cite{Weiss} can be easily shown to agree with the standard IF recast in so-called blip-sojourn form using sum and difference coordinates $\bar{s}(t)\equiv s^+(t)+s^-(t)$, $\Delta s(t)\equiv s^+(t)-s^-(t)$\cite{Makri2017}. The counting term derived in 
 \cite{Weiss}  is somewhat more involved and it reads 
\begin{widetext}
\begin{equation}
\begin{split} 
I^{(c)}=\exp\int_0^td\tau\int_0^\tau d\tau' \left\{\big[\bar{s}(\tau)\bar{s}(\tau')-\Delta s(\tau)\Delta s(\tau')\big]iL_1^\xi(\tau-\tau')+\big[\bar{s}(\tau)\Delta s(\tau')-\Delta s(\tau)\bar{s}(\tau')\big]iL^\xi_2(\tau-\tau')\right\},
\end{split}
\end{equation}
with the counting-dressed correlation functions given as
\begin{equation}
\begin{split}
	L_1^\xi(t)=&\frac{1}{\pi}\int_0^\infty d\omega g(\omega)\frac{\sinh(\beta\omega/2-i\omega\xi/2)}{\sinh(\beta\omega/2)}\cos(\omega t)\sin(-\omega\xi/2)\\
	L_2^\xi(t)=&-\frac{i}{\pi}\int_0^\infty d\omega g(\omega)\frac{\cosh(\beta\omega/2-i\omega\xi/2)}{\sinh(\beta\omega/2)}\sin(\omega t)\sin(-\omega\xi/2).
\end{split}
\end{equation}
Note that we flipped the sign of the counting field, $\xi$ from the original work to match our notation. This has the effect only of reversing 
the sign of the current, which is by convention taken positive when flowing towards the system.
We expand and multiply the sum and difference coordinates and retrieve the following,
\begin{equation}
	I^{(c)}=\exp\int_0^td\tau\int_0^\tau d\tau' \left[s^-(\tau)s^+(\tau')C_1^\xi(\tau-\tau')+s^+(\tau)s^-(\tau')C^\xi_2(\tau-\tau')\right].
\label{eq:Weiss_our_form}
\end{equation}
The new correlation functions are organized from the rearrangement of the coordinates, and come out as follows,
\begin{equation}
\begin{split}
	C_1^\xi(t)=&\frac{2}{\pi}\int_0^\infty d\omega g(\omega)\frac{\sin(-\omega\xi/2)}{\sinh(\beta\omega/2)}\left[\cosh(\beta\omega/2-i\omega\xi/2)\sin(\omega t)+i\sinh(\beta\omega/2-i\omega\xi/2)\cos(\omega t)\right]\\
	C_2^\xi(t)=-&\frac{2}{\pi}\int_0^\infty d\omega g(\omega)\frac{\sin(-\omega\xi/2)}{\sinh(\beta\omega/2)}\left[\cosh(\beta\omega/2-i\omega\xi/2)\sin(\omega t)-i\sinh(\beta\omega/2-i\omega\xi/2)\cos(\omega t)\right].
\end{split}
\end{equation}
These expressions are somewhat obtuse in their current forms,  and we decompose them into their real ($\Re$) and imaginary ($\Im$) parts, 
\begin{equation}
	\begin{split}
		\Re\big[C^\xi_1(t)\big]=&\frac{1}{\pi}\int_0^\infty d\omega g(\omega)\coth\bigg(\frac{\beta\omega}{2}\bigg)\big[\cos(\omega(t+\xi))-\cos(\omega t)\big]\\
		\Im\big[C^\xi_1(t)\big]=&-\frac{1}{\pi}\int_0^\infty d\omega g(\omega)\big[\sin(\omega(t+\xi))-\sin(\omega t)\big]\\
		\Re\big[C^\xi_2(t)\big]=&\frac{1}{\pi}\int_0^\infty d\omega g(\omega)\coth\bigg(\frac{\beta\omega}{2}\bigg)\big[\cos(\omega(t-\xi))-\cos(\omega t)\big]\\
		\Im\big[C^\xi_2(t)\big]=&\frac{1}{\pi}\int_0^\infty d\omega g(\omega)\big[\sin(\omega(t-\xi))-\sin(\omega t)\big].
	\end{split}
\end{equation}
Combining these terms, we re-express the correlation functions as
\begin{equation}
	\begin{split} 
		&C^\xi_1(t)=\frac{1}{\pi}\int_0^\infty d\omega g(\omega)\bigg[\coth\bigg(\frac{\beta\omega}{2}\bigg)
		\big[\cos(\omega(t+\xi))-\cos(\omega t)\big]-i\big[\sin(\omega(t+\xi))-\sin(\omega t)\bigg]\\
		&C^\xi_2(t)=\frac{1}{\pi}\int_0^\infty d\omega g(\omega)\bigg[\coth\bigg(\frac{\beta\omega}{2}\bigg)
		\big[\cos(\omega(t-\xi))-\cos(\omega t)\big]+i\big[\sin(\omega(t-\xi))-\sin(\omega t)\bigg].
	\label{eq:Weiss_IF}
\end{split}
\end{equation}
We are now ready to show that these expressions, adopted from Ref. \cite{Weiss}, agree with Eq. (\ref{eq:IF}) from the main body text.
First, we define $\eta^{(c)}(t+\xi)\equiv\eta(t+\xi)-\eta(t)$, excluding the normal kernel. 
It can be shown that  $\eta^{(c)}(\tau-\tau'+\xi)=C_1^\xi(\tau-\tau')$ and $\eta^{(c)}(\tau'-\tau+\xi)=C_2^\xi(\tau-\tau')$. To make this identification clear, we re-express the counting dressed continuous time correlation function, Eq. (\ref{eq:cont_corr}) as an integral from zero to infinity.
\bea
\eta^{(c)}(t+\xi)&=&\frac{1}{\pi}\int_0^\infty d\omega g(\omega)\left[\coth\left(\frac{\beta\omega}{2}\right)\cos(\omega (t+\xi))-i\sin(\omega (t+\xi))\right] - g(\omega)\left[\coth\left(\frac{\beta\omega}{2}\right)\cos(\omega t)-i\sin(\omega t)\right]
\nonumber\\
&=&\frac{1}{\pi}\int_0^\infty d\omega g(\omega)\left[\coth\left(\frac{\beta\omega}{2}\right)\left[\cos(\omega (t+\xi))-\cos(\omega t)\right]-i\left[\sin(\omega (t+\xi))-\sin(\omega t)\right]\right].
\eea
For the second correlation function, we have
\begin{equation}
\eta^{(c)}(-t+\xi)=\frac{1}{\pi}\int_0^\infty d\omega g(\omega)\left[\coth\left(\frac{\beta\omega}{2}\right)\left[\cos(\omega (t-\xi))-\cos(\omega t)\right]+i\left[\sin(\omega (t-\xi))-\sin(\omega t)\right]\right],
\end{equation}
thus $\eta^{(c)}(\tau'-\tau+\xi)=C_2^\xi(\tau-\tau')$. 
Altogether, when we use these functions in Eq. (\ref{eq:Weiss_our_form}) we get
\begin{equation}
{I}^{(c)}=\exp\int_0^t d\tau \int_0^\tau  d\tau' \left[s^-(\tau)s^+(\tau')\eta^{(c)}(\tau-\tau'+\xi)+
s^+(\tau)s^-(\tau')\eta^{(c)}(\tau'-\tau+\xi)\right],
\end{equation}
which corresponds to Eq. (\ref{eq:IF}) in the main text, stripped from the ``normal" influence functional.
\end{widetext}


\renewcommand{\theequation}{C\arabic{equation}}
\setcounter{equation}{0}  
\section*{Appendix C: Error analysis and symmetrization of the influence functional}


Iterative path integral schemes suffer from known numerical challenges,  
depending on the precise implementation and the system under study. 
For a recent example, Strathearn and coworkers \cite{Strathearn2018} 
found that for certain forms of the spectral density function, 
the path integral may produce nonphysical long-time behaviors. 
Here, we inherit the errors that come from the Trotter splitting of the propagator and the 
truncation of the memory kernel. Strategies to converge out these errors, as shown in e.g. Fig. \ref{fig:jvsg}, are generally successful here. 

\begin{figure}[htpb]
\includegraphics[scale=.45]{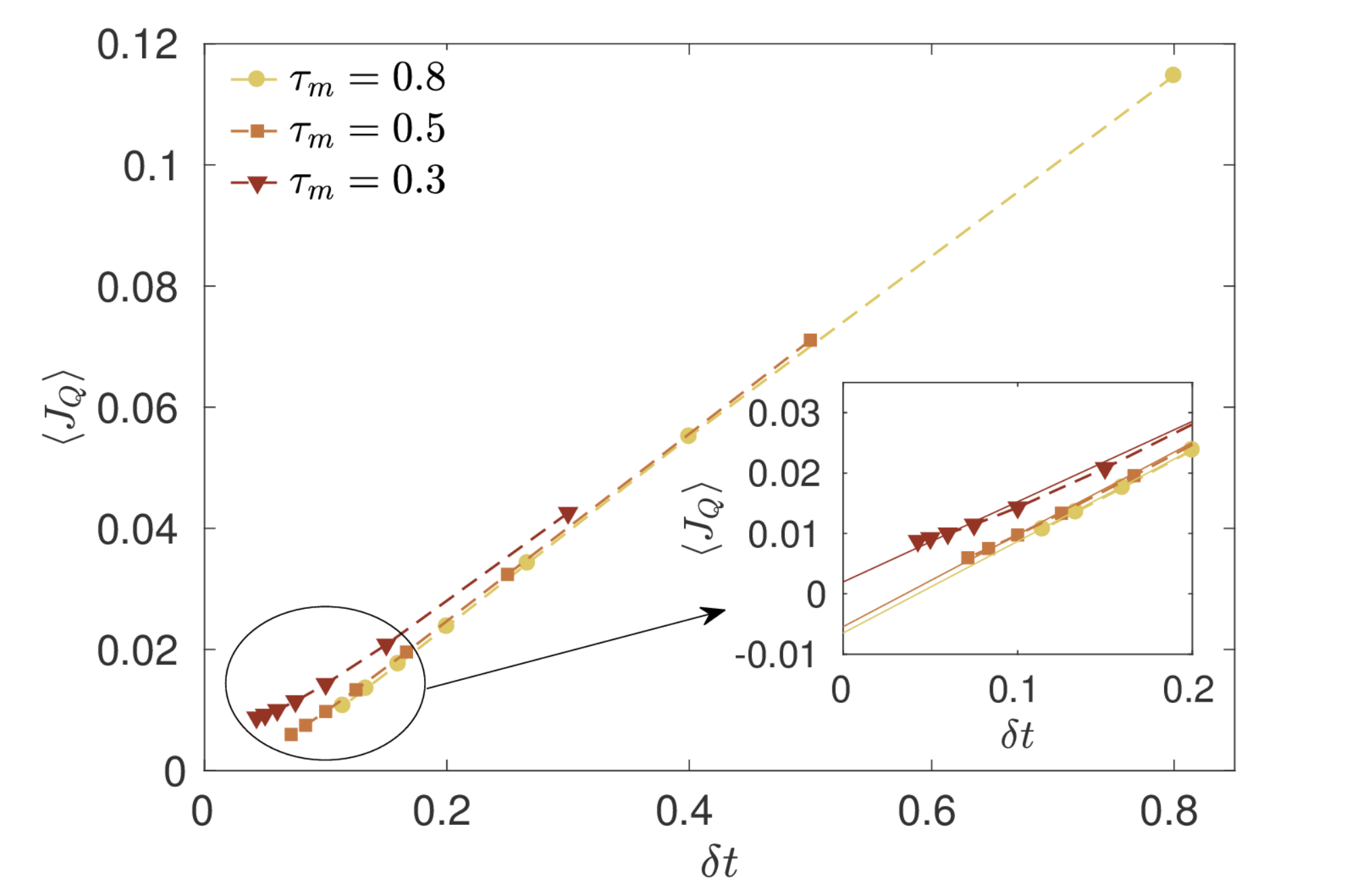} 
\caption{Error Analysis:
An asymmetric counting of energy exchange in the NESB model
leads to nonzero heat current at zero temperature difference.
We display a fixed-memory analysis of the nonphysical current by reducing the time step
as in Figure \ref{fig:jvsg}.
Inset: linear fit demonstrating the behavior of the spurious current at $\delta t\to 0$.
Parameters are $\Delta=1$, $T_{L}=T_R=5$, $\gamma_{L,R}=0.01$, $\omega_c=50$.
}
\label{fig:single_bath_counting}
\end{figure}

\begin{figure*}[htpb]
\includegraphics[scale=.6] {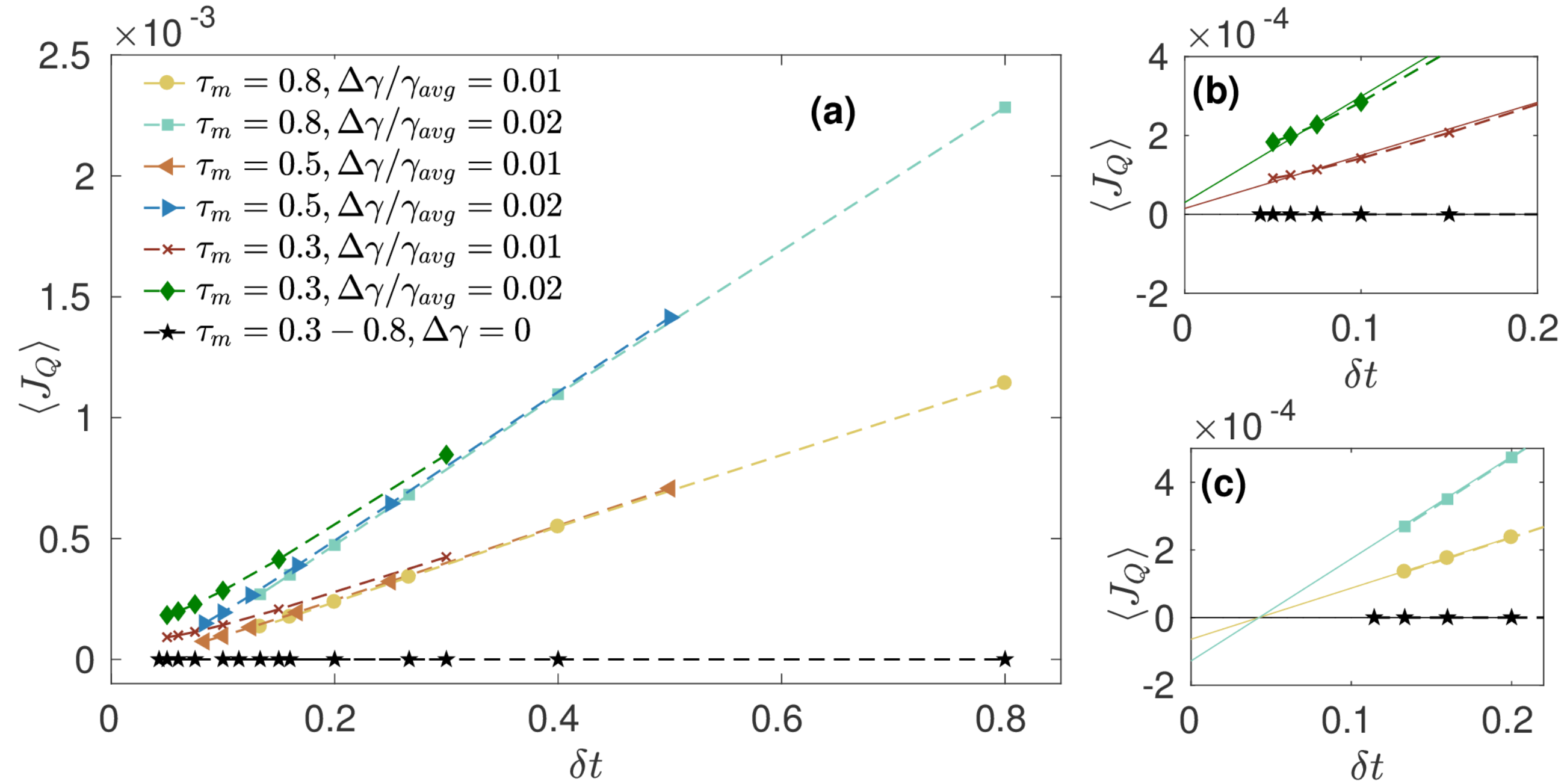} 
\caption{Error Analysis:
(a) Nonzero heat current at zero temperature difference when $\gamma_L\neq \gamma_R$.
The counting field is applied symmetrically to the $L$ and $R$ baths,
which are coupled asymmetrically to the spin system.
(b)-(c)  Linear fit of the spurious current, demonstrating its behavior as $\delta t\to 0$.
Parameters are $\Delta=1$, $T_{L}=T_R=5$, $\gamma_{avg}=0.01$, $\omega_c=50$.}
\label{fig:asym_error}
\end{figure*}

The familiar Trotter error manifests itself in new ways in the iFCSPI, in steady state. 
For example, when naively using a non-symmetric form of the heat current expression 
for a {\it single bath or multi-bath} setup
\bea
Q(t,0)=\hat{H}_L(0)-\hat{H}_L^H(t), 
\eea
one develops a significant erroneous current in the long time limit, 
appearing at zero temperature bias.
We exemplify this issue in Fig. \ref{fig:single_bath_counting}
where we display the iFCSPI steady state heat current in a junction at {\it zero} 
temperature difference ($T_L=T_R$), with the counting field dressing only the left bath. 
To understand the origin of this spurious effect, we perform a fixed-memory analysis so as 
to determine the scaling of the error with the time step $\delta t$.

We use a range of memory times ($\tau_m=\Delta k\cdot \delta t$) in Fig. \ref{fig:single_bath_counting}:
Simulations with $\tau_m=0.5, 0.8$ reasonably agree with each other, 
indicating that the total memory time is encapsulated in $\tau_m\sim 0.5$.
Most notably, we find that the nonphysical current diminishes
with $\delta t$, and that it linearly extrapolates to a value near, but not always exactly at zero for $\delta t=0$. 
The discrepancy from zero is usually small and depends on the parameters used.
It indicates that the scaling is not strictly linear for all parameter regimes at all values of 
$\delta t$. 
An important point is that the same erroneous phenomenon is observed even when the right bath is removed altogether. 

The nonphysical current demonstrated in Fig. \ref{fig:single_bath_counting}
can be quenched by reducing the time step $\delta t$ to the required accuracy. Naturally, this approach is
difficult to follow numerically, and we opt for a more economic approach, that is, symmetrizing the 
definition of heat exchange, as we discuss in Sec. \ref{Sec-NESB}, 
\bea
Q(t,0)=\frac{1}{2}\Big[\hat{H}_L(0)-\hat{H}_L^H(t)-\hat{H}_R(0)+\hat{H}_R^H(t)\Big].
\nonumber\\
\eea
This form results in strictly zero current at zero temperature difference for the two-bath case, when $g_L(\omega) =g_R(\omega)$.
We show this by taking a look at the discretized influence functional, Eq. (\ref{eq:disc_IF}).
When counting at both baths, we  consider the combination
$\eta^{+-,\xi}_{kk',L}+\eta^{+-,-\xi}_{kk',R}$.
Since
$\eta^{+-,-\xi}_{kk',R}=(\eta^{+-,\xi}_{kk',L})^*$ when $T_L=T_R$ and $g_L(\omega)=g_R(\omega)$, the sum of the kernels is an even function in 
$\xi$, thus the moment generating function provides identically zero odd moments of heat exchange, specifically zero averaged current. 

However, a related error (of finite current at zero temperature bias) arises whenever $g_L(\omega)\neq g_R(\omega)$,
which we demonstrate in Fig. \ref{fig:asym_error}. Here again we test the equilibrium scenario
and find nonzero current behaving as in Fig. \ref{fig:single_bath_counting}. 
The erroneous heat current 
scales linearly with the asymmetry in the system-bath coupling 
$\Delta\gamma/\gamma_{avg}$, with $\gamma_{avg}=(\gamma_L+\gamma_R)/2$ and $\Delta \gamma=\gamma_L-\gamma_R$.
We infer via the scaling with $\delta t$ that the error (nonzero current) is introduced by the 
Trotter splitting of the counting-dressed bath Hamiltonians, 
and that it disappears exactly at finite-large  $\delta t$ only under the choice of identical spectral functions for the baths with symmetric 
counting fields. 

The results presented in this main body of the paper were all 
computed from the symmetrized definition of the heat and using identical spectral functions for the reservoirs.
We emphasize that the iFCSPI method can be used beyond that---once significantly shorter time steps are employed.
The symmetrization of the counting field introduces a significant
limitation of the approach as yet, since the method cannot be straightforwardly employed beyond the case of two baths, e.g. to describe the behavior of
a (three bath) quantum absorption refrigerator.


A further challenge encountered with the current implementation of the 
iFCSPI is that the real part of the generating function is more difficult to converge
compared to the imaginary part. 
To illustrate the technique, we focused here only on the behavior of the averaged heat current.
It bears repeating however, that powerful new algorithms exist that should 
be able to tackle this issue.


\begin{thebibliography}{5}
\bibitem{thossC}
The heat current had to be symmetrized in other computational studies of
heat transport, see e.g. Refs. \cite{num-multi,ThossJCP}.
\end{thebibliography}
\end{document}